\documentclass[12pt]{article}
\usepackage{amsmath}
\usepackage{amssymb}

\usepackage{graphicx}
\usepackage{latexsym}
\usepackage{bm}
\usepackage{color}
\usepackage{enumitem}
\usepackage[toc,page]{appendix}
\usepackage{mathtools}
\usepackage{amsmath}
\usepackage{slashed}
\usepackage{hyperref}
\usepackage{tikz}
\usetikzlibrary{decorations.pathmorphing}
\numberwithin{equation}{section}

\tikzset{snake it/.style={decorate, decoration=snake}}

\begin{document}

\begin{titlepage}
    \setcounter{page}{1} \baselineskip=15.5pt \thispagestyle{empty}
    
    \begin{flushright}
		{\footnotesize{SLAC-PUB-16993, SU-ITP-17/07}} \date{}
	\end{flushright}
	
	\vspace{0.1cm}
	\begin{center}
		{\fontsize{21}{34}\selectfont \sc
		Heavy Fermion Production and \\[.3cm] Primordial $N$-Spectra
		\vspace{4mm}
		}
	\end{center}
	
	\begin{center}
		{\fontsize{13}{30}\selectfont Danjie Wenren}
	\end{center}
	
	\begin{center}

		\textsl{Stanford Institute for Theoretical Physics, Stanford University, Stanford, CA 94305, USA and}

		\textsl{ SLAC National Accelerator Laboratory, 2575 Sand Hill Rd., Menlo Park, CA 94025, USA}
\end{center}
\vspace{.5cm}
{
\begin{center}
\textbf{Abstract} 
\end{center}
\noindent We compute the non-adiabatic production of heavy fermion during inflation due to its coupling with inflaton. The coupling, partly inspired by axion monodromy, comes from the modulation of the fermion mass by the inflaton field. Even though the fermion mass is always much higher than the Hubble scale and the density of the produced fermions is low, they can still have detectable signatures in the cosmic microwave background. Their contributions to the primordial $N$-spectra are then analyzed in detail and compared with those from the fermion's bosonic super-partner.  At the classical level, where the produced particles are treated as classical sources, the effect on the $N$-spectra is proportional to the density of the produced particles and the fermion and boson cases have the same contribution. Quantum interference, however, leads to distinction between the two cases. Implications of this similarity and distinction are discussed before making general remarks about the limitations of our calculation and possible ways of overcoming them.
\vspace{0.3cm}
 }
	
\end{titlepage}

	\section{Introduction}
	The cosmic microwave background (CMB) observation and potentially the large-scale structure (LSS) observation in the future offer exciting opportunities to test high energy physics whose energy scale is far beyond what other types of experiments can achieve. In particular the oscillatory features of those observations can in principle constrain the possible ``ultra-violet complete'' theories of gravity. Such theories are needed in order to fully understand inflation, which is currently the most compelling theory framework of explaining the CMB data. With the significant amount of CMB data being collected by {\em Planck} \cite{Ade:2015ava, Ade:2015xua, Ade:2015lrj} it is worth studying from the theory perspective what physics mechanism can generate primordial seeds and providing searching templates for the corresponding signals.
	
	String theory as one of the most prominent candidates of theories of quantum gravity has offered a variety of ideas for inflation, of which axion monodromy \cite{Silverstein:2008sg, McAllister:2008hb, Flauger:2009ab} is a very natural one. Inspired in part by this mechanism, authors of \cite{Flauger:2016idt} consider theories where heavy boson fields interact with inflaton through non-derivative couplings. Such interaction comes from the fact that the boson mass is modulated by the inflaton field. As is well known in quantum field theory, such modulation can lead to non-adiabatic production of particles. It is found in \cite{Flauger:2016idt} that in a well-defined window of parameters as discussed there in, current CMB data is sensitive to such non-adiabatic production, even if the particle mass is always much higher than the Hubble scale $H$. The production's contribution to the primordial $N$-spectra is studied in detail and it is shown that the mechanism can generate oscillatory non-Gaussianity parametrically larger than that from the previously studied resonant non-Gaussianity.
	
	In this work we consider the effects of non-adiabatic production of heavy fermions which are super-partners of the bosons as in \cite{Flauger:2016idt}. As shown in the appendix of that paper, the radiative corrections to the inflaton $N$ point functions from these super-partners are suppressed if some degree of supersymmetry is assumed. Therefore in that case one only needs to consider the contributions from the particle production effects. It remains unclear, however, the relative scale between the contributions from boson production and those from fermion production, and if there can be novel feature shapes from fermion production. 
	
	In this work we will find that at the classical level, namely treating the produced particles as classical source, the two types of particles have the same contribution. The contribution due to quantum interference, however, is different between the fermion case and the boson case.
	
	The basic setup and particle mass as a function of inflaton field are reviewed in section \ref{section:setup}. In order to make a direct comparison with the boson results we work with the same mass function and in the same parameter regime as \cite{Flauger:2016idt}. The free fermion equation of motion in the inflation background is then solved using WKB approximation in section \ref{section:wkb:bogoliubov} along with the corresponding Bogoliubov transformations. Due to the relatively more complicated equation of motion and large minimum mass of the fermion field, a different method needs to be used than previous works in literature \cite{Peloso:2000hy, Chung:1999ve}. We found that in the parameter regime we consider, where the produced particle density is low and gets diluted rapidly by the exponential expansion of the universe, the Bogoliubov transformation is similar as in the boson case. In section \ref{section:powerspectrum:nongaussianity}, we compute two types of contributions to the inflaton $N$ point functions, one from treating the produced particle as classical source and the other from the quantum interference of the heavy particle fields. Comparisons between the fermion case and the boson case are also made in that section. Summary and a brief discussion of future directions are presented in section \ref{section:summary:discussion}.

	\section{Setup}
	\label{section:setup}
	It is discussed in \cite{Flauger:2016idt} that the radiative correction from heavy boson fields can be suppressed by the correction from their fermion partners assuming some degree of supersymmetry. As a result, the only contributions to the $N$-spectra will be those from particle production effects. In order for the radiative corrections to cancel, the boson mass $m_b$ and fermion mass $m_f$ need to satisfy $|m_b|^2 = |m_f|^2$ apart from the matching of numerical factors from particle and anti-particle doubling and helicity doubling. This does not, however, immediately determine $m_f$ as a function of the inflaton field $\phi$ since in a supersymmetric theory $m_f$ can be complex. We will work with an $m_f$ in the following that simplifies the analysis.
	
	Denoting the fermion field as $\psi$, its mass $M$ as a function of the inflaton field $\phi$ has two possible forms \cite{Flauger:2016idt}.
	\begin{enumerate}[label=(\alph*)]
		\item This is when the theory has multiple fields that undergo monodromy and each of them reaches its minimum mass when the inflaton traverses an underlying period in the axion field space. Let $n$ label different species of such fields and their masses can be written as 
			\begin{equation}
				|M|^2 = \mu_a^2 + \hat{\mu}_a^2[a(\phi) - 2\pi n]^2 \simeq \mu_a^2 + g_a^2(\phi - 2\pi n f),
			\end{equation}
			where we used $a(\phi) = \phi/f$ with $f$ being the decay constant and $g_a = \hat{\mu}_a /f$.
		\item In this case there is only one field whose mass is sinusoidally modulated by the inflaton field
			\begin{equation}
				\label{equation:mass:sqaure:cosine}
				|M|^2 = \mu^2 + 2g^2 f^2 \cos\frac{\phi}{f}.
			\end{equation}
			The positivity of $|M|^2$ requires 
			\begin{equation}
				\label{condition:M:positivity}
				gf < \frac{\mu}{\sqrt{2}}.			
			\end{equation}
			 We will use the WKB method to solve wave equations of the $\psi$ field and compute the number density of the produced $\psi$ particles. The production happens near the minimum mass region and for the $n$th such region, the mass can be expanded as 
			\begin{equation}
				|M|^2 \simeq \mu^2 -2g^2 f^2 + g^2(\phi - \phi_n)^2 = \mu_b^2 + g^2 \dot{\phi}^2 (t - t_n)^2,
			\end{equation}
			where $\phi_n = (2n+1 )\pi f$ and $\phi(t_n) = \phi_n$. In order for the WKB method to be valid, the production region should be within each oscillation period. The production region can be determined by maximizing $\dot{\omega}(p)/\omega^2(p)$ where $\omega(p)^2 = p^2 + |M|^2$ is the physical frequency of $\psi$ modes. A straightforward calculation shows that the maximum is at $|t - t_n| = \mu_b / (\sqrt{2}g\dot{\phi})$. Therefore the requirement of production happening well inside each oscillation period translates into the following condition
			\begin{equation}
				\label{condition:WKB:validity}
				\frac{\mu_b}{\sqrt{2}g\dot{\phi}} < \frac{\pi}{\dot{\phi}/f} \Rightarrow gf > \frac{\sqrt{2}\mu_b}{\pi},
			\end{equation}
			which can be consistent with condition (\ref{condition:M:positivity}).
	\end{enumerate}
	
	Notice that in both cases the mass has the same quadratic dependence on $\phi$ in the production region so we will first discuss their single production events and $N$ point correlation functions in a similar way. To explicitly evaluate the contributions of particle production events to the $N$ point functions we focus on case (b) since it provides novel searching templates while the other case overlaps strongly with existing ones.
	
	The action of the fermion field $\psi$ can then be written down as 
	\begin{equation}
		\label{action:fermion_1}
		S = \int d\eta d^3x a^4 \left[i\bar{\psi} ~e^{\mu}_{a} \gamma^a \nabla_{\mu} \psi - M(\phi) \bar{\psi} \psi \right],
	\end{equation}
	where the vierbein $e^a_{\mu}$ is the constant $4\times 4$ matrix $\frac{1}{a}I$ with $\nabla_{\mu}$ being the covariant derivative. Substituting their explicit forms in the FRW background into (\ref{action:fermion_1}) we get
	\begin{equation}
		\label{equation:action:psi}
		S = \int d\eta d^3x a^4\bar{\psi} \left[\frac{i}{a}\gamma^{\mu} \partial_{\mu} + i\frac{3}{2}H\gamma^0 - M\right]\psi,
	\end{equation}
	and the gamma matrices satisfy the usual anti-commutation relation $\{\gamma^{\mu}, \gamma^{\nu}\} = -2 \eta^{\mu\nu}$ where $\eta_{\mu\nu}$ is the metric for flat spacetime $\rm{diag}(-1, +1, +1, +1)$ \footnote{Note that strictly speaking, $\gamma^{\mu}\partial_{\mu}$ should be understood as $\gamma^{a} I_a^{\mu} \partial_{\mu}$ and $\gamma^{a}$ are the matrices that satisfy the usual anti-commutation relation in flat spacetime. For notational simplicity we will ignore this detail in this work.}. Defining $\psi = a^{-3/2}\chi$ the action simplifies to 
	\begin{equation}
		\label{equation:action:chi}
		S = \int d\eta d^3 x \bar{\chi} \left[i\gamma^{\mu} \partial_{\mu} - aM\right]\chi.
	\end{equation}
	As in \cite{Flauger:2016idt} we consider the parameter regime where the heavy particles do not back-react on the inflation background $\phi_0$ and the $\chi$ equation of motion is solved in this background. More explicitly we will work with the following free and interaction Lagrangians
	\begin{align}
		\label{equation:free:lag:chi}
		\mathcal{L}_0 &= \bar{\chi}\left[i\gamma^{\mu} \partial_{\mu} - aM(\phi_0)\right]\chi \\%
		\label{equation:interaction:lag:chi}
		\mathcal{L}_I &= -a\frac{\delta M}{\delta\phi}\Big|_{\phi_0}\delta\phi\bar{\chi}\chi - \frac{1}{2}a\frac{\delta^2 M}{\delta\phi^2}\Big|_{\phi_0}\delta\phi^2\bar{\chi}\chi - \cdots.
	\end{align}
	
	\section{WKB solution for fermion field and Bogoliubov transformation}
	\label{section:wkb:bogoliubov}
	The free equation of motion of $\chi$ can be derive from (\ref{equation:free:lag:chi}) as 
	\begin{equation}
		\label{equation:chi:eom}
		(i\gamma^{\mu} \partial_{\mu} - aM)\chi = 0.
	\end{equation}
	It can be solved using Fourier transformation. To begin with, let us write $\chi(\eta, \bf{x})$ as
	\begin{equation}
		\label{equation:chi:expansion}
		\chi = \int_{\bf{k}} e^{i\bf{k} \cdot \bf{x}}\sum_r \left[u_r(\eta, {\bf k}) a_r({\bf k}) + v_r(\eta, {\bf k}) b_r^{\dagger}(-\bf{k})\right],
	\end{equation}
	where $\int_{\bf{k}}$ is a short-hand notation for $\int\frac{d^3 \bf{k}}{(2\pi)^3}$ and $r = \pm 1$ denotes the helicity of the $\chi$ modes. The mode functions can be further decomposed as 
	\begin{equation}
		u_r = \frac{1}{\sqrt{2}}
			\begin{pmatrix}
				u_+(\eta, {\bf k}) \psi_r({\bf k})\\
				u_-(\eta, {\bf k})\psi_r({\bf k})
			\end{pmatrix},
		\quad
		v_r = \frac{1}{\sqrt{2}}
			\begin{pmatrix}
				v_+(\eta, {\bf k}) \psi_r({\bf k})\\
				v_-(\eta, {\bf k}) \psi_r({\bf k})
			\end{pmatrix}
	\end{equation}
	where the two-component vectors $\psi_r$ are helicity eigenstates satisfying ${\bf k}\cdot {\bf \sigma}\psi_r = rk\psi_r$ with $\sigma^i$ being the Pauli matrices. 
	
	Before we substitute (\ref{equation:chi:expansion}) into  (\ref{equation:chi:eom}) we also need to select a basis for the gamma matrices. While in principle different bases should lead to the same physical result, we will use the Dirac basis where 
	\begin{equation}
		\gamma^0 = 
			\begin{pmatrix}
				I & 0\\
				0 & -I
			\end{pmatrix},~
		\gamma^i = 
			\begin{pmatrix}
				0 & \sigma^i\\
				-\sigma^i & 0
			\end{pmatrix},~
		\gamma^5 = 
			\begin{pmatrix}
				0 & I\\
				I & 0
			\end{pmatrix}.
	\end{equation}
	It works better than the other commonly used one, namely the Weyl basis where the $\gamma^i$ matrices are the same while the other two are
	\begin{equation}
		\gamma^0 = 
			\begin{pmatrix}
				0 & I\\
				I & 0
			\end{pmatrix},
		\qquad
		\gamma^5 = 
			\begin{pmatrix}
				-I & 0\\
				0 & I
			\end{pmatrix}.
	\end{equation}
	This is because of the more complicated wave equations the latter leads to. We will first derive the wave equations for $u_{\pm}$ in Dirac basis and then demonstrate why it is not as convenient in Weyl basis. The mode functions $v_{\pm}$ are related to $u_{\pm}$ as 
	\begin{equation}
		\label{equation:relation:uv}
		v_+ = -u_-^*, \qquad v_- = u_+^*,
	\end{equation}
	so we will not explicitly write out the wave equations for $v_{\pm}$ \cite{Peloso:2000hy}.
	
	\subsection{Wave equation in Dirac basis}
	\label{sec:Dirac:basis}
	Using (\ref{equation:chi:eom}) and other definitions above, one can get the equations of motion for $u_{\pm}$ as 
	\begin{align}
		\label{equation:u:eom:1order}
		iu_+' + rku_- -aMu_+ = 0, \quad iu_-' + rku_+ + aMu_- = 0,
	\end{align}
	where prime means taking derivative with respect to $\eta$. These first order equations can then be combined into second order equations similar to the Klein-Gordon equation
	\begin{equation}
		\label{equation:u:eom:2order}
		u_{\pm}'' + \left[k^2 + a^2 M^2 \pm i (aM)'\right]u_{\pm} = 0.
	\end{equation}
	This form is relatively simple and we will solve it using WKB approximation.
	
	One can also easily verify using (\ref{equation:u:eom:1order}) that $(|u_+|^2 + |u_-|^2)' = 0$ so we will normalize the mode functions as $|u_+|^2 + |u_-|^2 = 2$, which can be used as a consistency check for the Bogoliubov coefficients we get. Note that in deriving this normalization condition we used the fact that $M$ is real. In a supersymmetric theory this is not always the case so we will need to be careful about the definition of $M$ in the following. We will prescribe a particular analytic structure for $M(\eta)$ in section \ref{subsection:wkb:bogo} which ensures that it is real on the real $\eta$ axis.
	
	\subsection{Wave equation in Weyl basis}
	In this basis the first order wave equations become
	\begin{equation}
		iu_-' + rku_- - aM u_+ = 0, \quad iu_+' - rku_+ - aMu_- = 0.
	\end{equation}
	Besides the problem of now having the mode functions depending on helicity $r$, we also get more complicated second order equations of motion
	\begin{equation}
		u_{\pm}'' + (k^2 + a^2 M^2) u_{\pm} - \frac{(aM)'}{aM}(u_{\pm}' \pm ir ku_{\pm}) = 0.
	\end{equation}
	Since this looks much less tractable than the Dirac basis case (\ref{equation:u:eom:2order}), the rest of our calculation will be in Dirac basis.
	
	\subsection{Solution in the WKB approximation} \label{subsection:wkb:bogo}
	In both cases (a) and (b) we can write the mass near the $n$th production region as 
	\begin{equation}
		\label{equation:mass:general}
		|M(\phi)|^2 = \tilde{\mu}^2 + g^2 (\phi - \phi_n)^2 = \tilde{\mu}^2 + g^2 a_n^2 \dot{\phi}^2 (\eta - \eta_n)^2,
	\end{equation}
	where $\phi = \phi_n$ is the point where $\psi$ reaches its minimum mass and dot represents the derivative with respect to physical time $t$. We have also used the relation $d/d\eta = a(d/dt)$. In section \ref{sec:Dirac:basis} we have restricted $M$ to be on the real axis but we still need to fix a branch cut when we take the square root of (\ref{equation:mass:general}). This is particularly an issue in the analysis here due to the imaginary term in (\ref{equation:u:eom:2order}). This can be better seen if we plug (\ref{equation:mass:general}) in and expand the derivative term
	\begin{equation}
		\label{equation:wave:expanded}
		u_+'' + \left(k^2 + a_n^2 \tilde{\mu}^2 + g^2 a_n^4\dot{\phi}^2 (\eta -\eta_n)^2 +
			ia_n \frac{g^2 a_n^2 \dot{\phi}^2 (\eta - \eta_n)}{\sqrt{\tilde{\mu}^2 + g^2 a_n^2 \dot{\phi}^2 (\eta -\eta_n)^2}}
			\right)u_+ = 0
	\end{equation}
	Note that since within Hubble the expansion is relatively slower than the changing in mass, we have treated $a$ as a constant $a_n$. Also for notational simplicity in this section we will make the substitution $\Delta\eta = \eta - \eta_n \to \eta$.
	
	A typical way of using WKB method to solve (\ref{equation:wave:expanded}) involves getting the exact solution of it and then matching it with the WKB solutions in the $\eta\to -\infty$ limit and the $\eta\to +\infty$ limit\footnote{Note that we have made the substitution $\eta - \eta_n \to \eta$ for notational simplicity.}  \cite{Peloso:2000hy, Chung:1999ve, Kofman:2004yc}. However unlike the wave equations in those works, it seems rather intractable to exactly solve the wave equation (\ref{equation:wave:expanded}) due to the inverse square root term. A simplification one might consider is that since the matching happens in the $|\eta| \to \infty$ region, the wave equation can be simplified in that region. This indeed happens because in that region the imaginary term becomes $ig a_n^2 |\dot{\phi}|$ in the $\eta\to +\infty$ limit and $-ig a_n^2 |\dot{\phi}|$ in the opposite limit. However this also shows that the solutions to (\ref{equation:wave:expanded}) cannot be extended to analytic functions on the full $\eta$ plane including $\infty$ since the solution should satisfy different differential equations at $-\infty$ and $+\infty$. We explore this method in the Appendix \ref{appendix:sec:bogo} and argue that it does not work due to the complicated Stokes phenomenon of the original wave equation (\ref{equation:wave:expanded}). In this section we instead use a different approach which directly rotates one WKB solution in one limit to the one in the other limit along a semi-circle on the complex $\eta$ plane.
	
		The WKB solution to (\ref{equation:wave:expanded}) takes the form \footnote{ Note that in our solution the definition for $A$ and $B$ is different from that in \cite{Peloso:2000hy} because of a sign difference in the free Hamiltonian (\ref{equation:hamiltonian:chi}).}
	\begin{align}
		\label{equation:WKB:uplus}
		u_+(\eta) &= A\left(1 + \frac{aM}{\omega}\right)^{\frac{1}{2}} e^{-i\int^{\eta}_0 d\eta' ~\omega(\eta')} + B \left(1 -\frac{aM}{\omega}\right)^{\frac{1}{2}} e^{i\int^{\eta}_0 d\eta'~\omega(\eta)} ,\\
		\label{equation:WKB:uminus}
		u_-(\eta) &= B\left(1+\frac{aM}{\omega}\right)^{\frac{1}{2}}e^{i\int^{\eta}_0 d\eta' ~\omega(\eta')} - A\left(1 - \frac{aM}{\omega}\right)^{\frac{1}{2}} e^{-i\int^{\eta}_0 d\eta' ~ \omega(\eta')},
	\end{align}
	with $A$ and $B$ being constants and $\omega^2 = k^2 + a_n^2 \tilde{\mu}^2 + g^2 a_n^4 \dot{\phi}^2 \eta^2$ the Fourier frequency\footnote{Note again that the lower limit $0$ in the integral means the production point since we have made the substitution $\eta - \eta_n \to \eta$ for the purpose of having less cumbersome equations.}. As usual we would have different $A$'s and $B$'s in the $\eta > 0$ region and $\eta < 0$ region due to the non-adiabaticity in the production region. In general one would need a recurrence relation for $A_{n}, B_{n}$ after the $n$th production event in terms of $A_{n-1}, B_{n-1}$ before the $n$th production event. However our method will only be able to give result for $B_{n}$ in the region where  $|B_{n-1}| \simeq 0$. As argued in \cite{Flauger:2016idt} this is actually good enough for our purpose since during each production event only a small amount of particles get produced and they are quickly diluted by inflation. We will see later that the density of the produced fermions is proportional to $|B|^2$ so we can always assume $|B|$ to be small. Moreover, the normalization condition $|u_+|^2 + |u_-|^2 = 2$ enforces $|A|^2 + |B|^2 = 1$ , therefore for each production event, it would be sufficient to start with only positive frequency modes, that is $A = 1$ and $B = 0$ in the $\eta \to -\infty$ limit and then solve for $B$ in the opposite limit. Without loss of generality, we will only make the rotation explicitly for $u_+$ since $u_-$ has the same set of parameters.
	
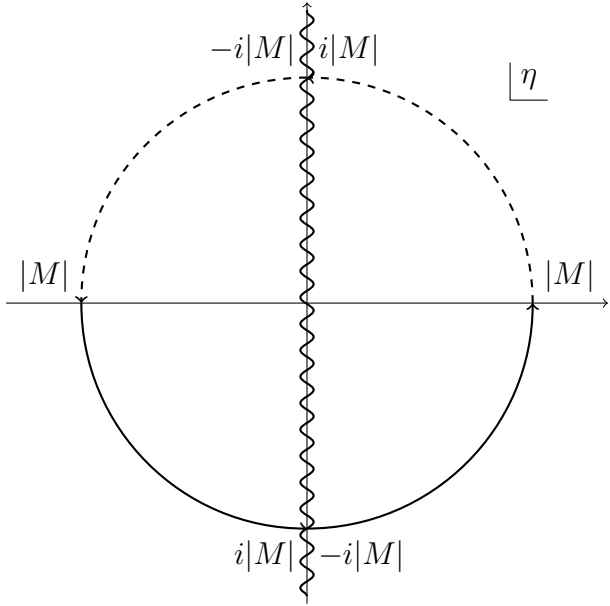
\begin{figure}
\centering
\begin{tikzpicture}
	\draw[->] (-4, 0) -- (4, 0);
	\draw[->] (0, -4) -- (0, 4);
	\draw[thick, dashed, ->] (3, 0) node[above right]{$|M|$} to [out=90, in=0] (0, 3) node[above right]{$i|M|$};
	\draw[thick, dashed, ->] (0, 3) node[above left]{$-i|M|$} to [out=180, in=90] (-3, 0) node[above left]{$|M|$};
	\draw[thick, ->] (-3, 0) to [out=270, in=180] (0, -3) node[below left]{$i|M|$};
	\draw[thick, ->] (0, -3) node[below right]{$-i|M|$} to [out=0, in=270](3, 0);
	\draw[thick, snake it] (0, -3.9) -- (0, 3.9);
	\draw[-] (2.7, 3.2) -- (2.7, 2.7) node[above right]{$\eta$} -- (3.2, 2.7);	
\end{tikzpicture}
\caption{The branch cut (wavy line) is on the the imaginary axis therefore the mass function changes sign in the middle of the rotation required for matching the two WKB solutions in the $\eta\to-\infty$ and $\eta\to+\infty$ limits, whose path is denoted by the solid semi-circle.}
\label{fig:mass:phase:diagram:wrong}
\end{figure}

	Defining $\nu_0^2 = k^2 + a_n^2 \tilde{\mu}^2$ and $\nu_1^4 = g^2 a_n^4 \dot{\phi}^2$ we can immediately write down the following two approximations needed for computing the WKB solutions, namely the phase
	\begin{equation}
		\label{equation:WKB:phase}
		\Phi(\eta) = \int^{\eta}_0 d\eta' \sqrt{\nu_0^2 + \nu_1^4 \eta'^2} \simeq
		\begin{cases}
			\frac{1}{2}\nu_1^2\eta^2 + \frac{\nu_0^2}{2\nu_1^2}\log\left(\frac{2\nu_1^2}{\nu_0}\eta\right) + \frac{1}{4}\frac{\nu_0^2}{\nu_1^2}, & \eta \to +\infty \\%
			-\frac{1}{2}\nu_1^2\eta^2 - \frac{\nu_0^2}{2\nu_1^2}\log\left(\frac{2\nu_1^2}{\nu_0}|\eta|\right) - \frac{1}{4}\frac{\nu_0^2}{\nu_1^2}, & \eta \to -\infty
		\end{cases}.
	\end{equation}
	and frequency
		\begin{equation}
		\label{equation:freq:expansion}
		\omega \simeq 
		\begin{cases}
			-\nu_1^2\eta\left(1 + \frac{\nu_0^2}{2\nu_1^4 \eta^2}\right), & \eta \to -\infty \\%
			\nu_1^2\eta\left(1 + \frac{\nu_0^2}{2\nu_1^4 \eta^2}\right), & \eta \to +\infty 		
		\end{cases}.
	\end{equation}
	The other one in need is the mass $M$. One needs to be more careful about it because if the usual branch cut is used, the rotation would cross the branch cut at $\arg \eta = \pm\frac{\pi}{2}$ as shown in Fig \ref{fig:mass:phase:diagram:wrong}. Because $M$ changes sign at the branch cut, there is no way to keep rotating the same analytic function to the other side of the real axis. 
	To show this more precisely, we can use the following mass function approximation
	\begin{equation}
		\label{equation:mass:expansion:wrong:bc}
		a_n M \simeq 
		\begin{cases}
			-\nu_1^2 \eta \left(1 + \frac{\nu_0^2 - k^2}{2\nu_1^4\eta^2}\right), & \eta\to - \infty \\%
			\nu_1^2 \eta \left(1 + \frac{\nu_0^2 - k^2}{2\nu_1^4\eta^2}\right), & \eta \to +\infty
		\end{cases}
	\end{equation}
	and spell out the WKB solution
	\begin{align}
		\eta \to -\infty, ~& u_+(\eta) \simeq \sqrt{2}e^{\frac{i\nu_0^2}{4\nu_1^2}}\left(\frac{2\nu_1}{\nu_0}\right)^{\frac{i\nu_0^2}{2\nu_1^2}}(-x)^{\frac{i\nu_0^2}{2\nu_1^2}}e^{\frac{i}{2}x^2} ,		\\%
		\eta \to +\infty, ~& u_+(\eta) \simeq \sqrt{2}A e^{-\frac{i\nu_0^2}{4\nu_1^2}}\left(\frac{2\nu_1}{\nu_0}\right)^{-\frac{i\nu_0^2}{2\nu_1^2}}x^{-\frac{i\nu_0^2}{2\nu_1^2}}e^{-\frac{i}{2}x^2} + 
				B\frac{k}{\sqrt{2}\nu_1}e^{\frac{i\nu_0^2}{4\nu_1^2}}\left(\frac{2\nu_1}{\nu_0}\right)^{\frac{i\nu_0^2}{2\nu_1^2}}x^{-1 + \frac{i\nu_0^2}{2\nu_1^2}}e^{\frac{i}{2}x^2},
	\end{align}
	where we have defined $x = \nu_1 \eta$. It is clear that one cannot rotate the first solution to the second one\footnote{Notice that neither of the terms in the $\eta\to +\infty$ limit matches the $x$ dependence in the $\eta\to -\infty$ limit.} due to the branch cut along $\arg \eta = -\frac{\pi}{2}$. 
	
\begin{figure}
\centering
\begin{tikzpicture}
	\draw[->] (-4, 0) -- (4, 0);
	\draw[->] (0, -4) -- (0, 4);
	\draw[thick, dashed, ->] (3, 0) node[above right]{$|M|$} to [out=90, in=0] (0, 3) node[above right]{$i|M|$};
	\draw[thick, dashed, ->] (0, 3) node[above left]{$i|M|$} to [out=180, in=90] (-3, 0) node[above left]{$-|M|$};
	\draw[thick, ->] (-3, 0) node[below left]{$|M|$} to [out=270, in=180] (0, -3) node[below left]{$i|M|$};
	\draw[thick, ->] (0, -3) node[below right]{$i|M|$} to [out=0, in=270](3, 0) node[below right]{$-|M|$};
	\draw[thick, snake it] (-3.9, 0) -- (3.9, 0);
	\draw[-] (2.7, 3.2) -- (2.7, 2.7) node[above right]{$\eta$} -- (3.2, 2.7);	
\end{tikzpicture}
\caption{The branch cut we use for matching the two WKB solutions. The matching rotates the solution in the $\eta\to-\infty$ limit to the one in the $\eta\to+\infty$ limit along a semi-circle on the lower half of the $\eta$ plane. With the branch cut shown here the rotation stays on one branch of the mass function except for the final infinitesimal part that crosses the real axis.}
\label{fig:mass:phase:diagram:right}
\end{figure}
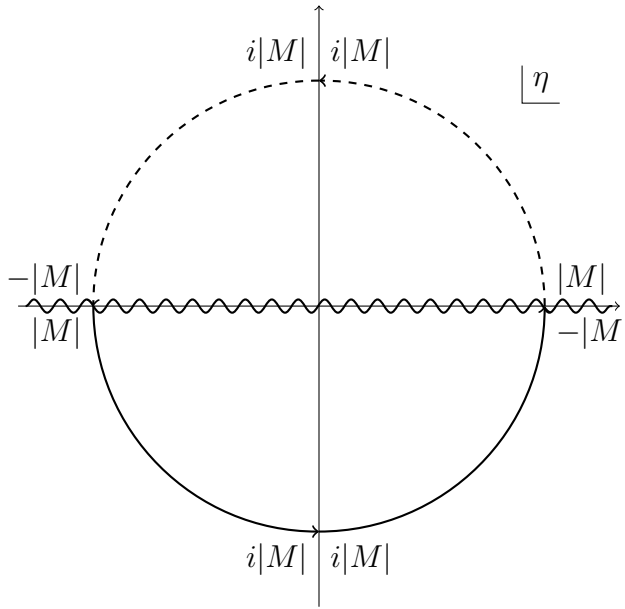

	To solve this problem a different branch cut needs to be used for the mass function, which is shown in Fig \ref{fig:mass:phase:diagram:right}. Essentially the branch cut of the square root function is taken from the negative real axis to the positive axis, so that the branch cuts on the $\eta$ plane are taken to the real axis. With this definition of branch cut, one could rotate the WKB solution in the $\eta\to-\infty$ region along a semi-circle to the $\eta\to +\infty$ region, where at both ends of the path the mass is positive. One caveat of this rotation is that in order to keep the mass positive in the end, one would still need to cross the branch from $+\infty(1-i\epsilon)$ to $+\infty(1 + i\epsilon)$, where $\epsilon >0$ is an arbitrarily small number. We expect, however, this infinitesimal rotation would not change the Bogoliubov coefficients in a significant way.
	With this definition of the mass function, we can write down the WKB solutions at the two ends of the rotation
	\begin{align}
		\label{equation:right:wkb:neg} \eta \to -\infty, ~& u_+(\eta) \simeq \sqrt{2}e^{\frac{i\nu_0^2}{4\nu_1^2}}\left(\frac{2\nu_1}{\nu_0}\right)^{\frac{i\nu_0^2}{2\nu_1^2}}(-x)^{\frac{i\nu_0^2}{2\nu_1^2}}e^{\frac{i}{2}x^2} ,		\\%
		\label{equation:right:wkb:pos}\eta \to +\infty, ~& u_+(\eta) \simeq A\frac{k}{\sqrt{2}\nu_1}e^{-\frac{i\nu_0^2}{4\nu_1^2}}\left(\frac{2\nu_1}{\nu_0}\right)^{-\frac{i\nu_0^2}{2\nu_1^2}}x^{-1-\frac{i\nu_0^2}{2\nu_1^2}}e^{-\frac{i}{2}x^2} + B\sqrt{2}e^{\frac{i\nu_0^2}{4\nu_1^2}}\left(\frac{2\nu_1}{\nu_0}\right)^{\frac{i\nu_0^2}{2\nu_1^2}}x^{\frac{i\nu_0^2}{2\nu_1^2}}e^{\frac{i}{2}x^2}
	\end{align}
	where the second limit is taken to be on the lower side of the real axis. It is now clear that (\ref{equation:right:wkb:neg}) rotates into the second term of (\ref{equation:right:wkb:pos}). To make this rotation, let $x = \rho e^{i\phi}$ and $\phi$ goes from $-\pi + \epsilon$ to $-\epsilon$ where $\epsilon > 0$ is an arbitrarily small number. The $(-x)^{\frac{i\nu_0^2}{2\nu_1}}$ factor can be written as $\left(e^{i\pi}e^{-i\pi}\rho\right)^{\frac{i\nu_0^2}{2\nu_1}}$ before rotation and becomes $\left(e^{i\pi}\rho\right)^{\frac{i\nu_0^2}{2\nu_1}}$ afterwards. This should match with the second term in (\ref{equation:right:wkb:pos}) and therefore yields $B = e^{-\frac{\pi\nu_0^2}{2\nu_1^2}}$. 
	
	One might worry about the appearance of the first term in (\ref{equation:right:wkb:pos}). To address this we notice that during the rotation, $e^{\pm \frac{i}{2}x^2} = e^{\frac{\rho^2}{2}(\mp\sin2\phi +\pm i\cos2\phi)}$. If one starts with (\ref{equation:right:wkb:pos}) and rotates it to (\ref{equation:right:wkb:neg}), the first term will be exponentially suppressed compared to the second one and therefore should not be kept in this WKB approximation \cite{landau:qm}. In other words, at the level of WKB approximation, (\ref{equation:right:wkb:neg}) and (\ref{equation:right:wkb:pos}) are the same solution and we cannot fix $A$ using this rotation method. If one is able to solve (\ref{equation:wave:expanded}) exactly then in principle $A$ can be computed. This possibility is explored in Appendix \ref{appendix:sec:bogo}.

	\subsection{Bogoliubov transformation}
	After each production event the creation and annihilation operators need to be redefined in order to diagonalize the Hamiltonian with a new set of mode functions $u_+$ and $u_-$. This can fix the Bogoliubov transformation
	\begin{align}
		\label{equation:bogoliubov:1}
		\hat{a}_ {\bf k}(\eta) = \alpha_{\bf k} a_{\bf k} + \beta_{\bf k}b^{\dagger}_{-\bf k}, \quad
		\hat{b}^{\dagger}_{-\bf k}(\eta) = -\beta^*_{\bf k} a_{\bf k} + \alpha^*_{\bf k} b^{\dagger}_{-\bf k}
	\end{align}
	in terms of $A$ and $B$. Note that the anti-commutation relation $\{\hat{a}_{\bf k}, \hat{a}^{\dagger}_{\bf k'}\}$ and $\{\hat{b}_{\bf k}, \hat{b}^{\dagger}_{\bf k'}\}$ immediately lead to the normalization condition $|\alpha_{\bf k}|^2 + |\beta_{\bf k}|^2 = 1$. In this subsection we will first solve for $\alpha$ and $\beta$ by explicitly diagonalizing the Hamiltonian and then show that it is equivalent to rewriting the $\chi$ expansion (\ref{equation:chi:expansion}) in terms of purely positive modes and negative modes.
	
	The free Hamiltonian can be derived from (\ref{equation:free:lag:chi}) as\footnote{Despite the fact that it is straightforward to compute the Hamiltonian, we still find a sign difference as compared with \cite{Peloso:2000hy}. It seems our result agrees with most textbooks. As a result our result (\ref{equation:EF:uplusminus}) also has a sign difference with that of  \cite{Peloso:2000hy}.} 
	\begin{equation}
		\label{equation:hamiltonian:chi}
		H_0 = \int d^3x~ i\bar{\chi}\gamma^0 \partial_{\eta}\chi = \int d^3x~ i \chi^{\dagger}\partial_{\eta}\chi,
	\end{equation}
	where we have used the free equation of motion of $\chi$ so that $H_0$ has only one time derivative term. Using the fermion field expansion (\ref{equation:chi:expansion}) it can be expressed in terms of $a$ and $b$ as
	\begin{equation}
		\label{equation:hamiltonian:EF}
		H_0 = \int_{\bf k} \sum_r \left[E({\bf k})\left(a_r^{\dagger}({\bf k})a_r({\bf k}) - b_r(-{\bf k}) b_r^{\dagger}(-{\bf k})\right)
			 +  F({\bf k}) b_r(-{\bf k}) a_r({\bf k})
			 + F^*({\bf k})a^{\dagger}_r({\bf k}) b^{\dagger}_r(-{\bf k})\right],
	\end{equation}
	where 
	\begin{align}
		\label{equation:EF:uplusminus}
		E &= \frac{1}{2}\left[aM\left(|u_+|^2 - |u_-|^2\right) - k\left(u_+^* u_- + u_+ u_-^*\right)\right] \nonumber \\%
		F &= \frac{1}{2}\left[k\left(u_-^2 - u_+^2\right) - 2aMu_+ u_-\right]
	\end{align}
	and $E^2 + |F|^2 = \omega^2$. 
	Using the reverse of (\ref{equation:bogoliubov:1}), that is,
	\begin{align}
		\label{equaton:bogoliubov:2}
		a_{\bf k} = \alpha^*_{\bf k} \hat{a}_{\bf k} - \beta_{\bf k}\hat{b}^{\dagger}_{-\bf k}, \quad
		b^{\dagger}_{- \bf k} = \beta^*_{\bf k} \hat{a}_{\bf k} + \alpha_{\bf k} \hat{b}^{\dagger}_{-\bf k},
	\end{align}
	and enforcing the vanishing of off diagonal terms (terms proportional to $\hat{b}\hat{a}$ and $\hat{a}^{\dagger}\hat{b}^{\dagger}$) lead to the following relations
	\begin{equation}
		\label{equation:bogo:inEF}
		\frac{\alpha}{\beta} = \frac{E + \omega}{F^*}, \quad |\beta|^2 = \frac{\omega - E}{2\omega}.
	\end{equation}
	In order to relate $\alpha$ and $\beta$ to $A$ and $B$ one can plug in the WKB solution (\ref{equation:WKB:uplus}) and (\ref{equation:WKB:uminus}) and get $E = \omega(|A|^2 - |B|^2)$ and $F = -2\omega AB$. So (\ref{equation:bogo:inEF}) gives $\alpha/\beta = -A/B^*$ and $|\beta|^2 = |B|^2$. This means that the diagonalization of the Hamiltonian does not completely fix $\alpha$ and $\beta$.  We pick $\alpha = A$ and $\beta = -B^*$ for the requirement that $\chi$ can be written as a sum of purely positive and negative modes in terms of $\hat{a}$ and $\hat{b}$ as shown in the following.
	
	Denote $\hat{u}_{\pm}$ and $\hat{v}_{\pm}$ as purely positive and negative modes, respectively then
	\begin{equation}
		\label{equation:definition:u:pm}
		\begin{cases}
			&\hat{u}_+ = \left(1 + \frac{aM}{\omega}\right)^{\frac{1}{2}} e^{-i\int \omega}\\%
			&\hat{u}_- = -\left(1-\frac{aM}{\omega}\right)^{\frac{1}{2}}e^{-i \int\omega}
		\end{cases}
		\qquad
		\begin{cases}
			&\hat{v}_+ = \left(1 - \frac{aM}{\omega}\right)^{\frac{1}{2}}e^{i\int \omega}\\%
			&\hat{v}_- = \left(1 + \frac{aM}{\omega}\right)^{\frac{1}{2}}e^{i\int\omega}
		\end{cases},
	\end{equation}
	where $\int \omega$ is a short hand for $\int_0^{\eta} d\eta'\omega(\eta')$. Then the WKB solution (\ref{equation:WKB:uplus}), (\ref{equation:WKB:uminus}) can be written as 
	\begin{equation}
		\begin{cases}
			&u_+ = A\hat{u}_+ + B\hat{v}_+ \\%
			&u_- = B\hat{v}_- + A\hat{u}_-
		\end{cases}
		\qquad
		\begin{cases}
			&v_+ = -u_-^* = -B^* \hat{u}_+ + A^*\hat{v}_+\\%
			&v_- = u_+^* = A^* \hat{v}_- - B^* \hat{u}_-
		\end{cases}.
	\end{equation}
	With this relation the integrand of the $\chi$ expansion (\ref{equation:chi:expansion}) becomes
	\begin{equation}
		\label{equation:chi:pure:positive}
		u_r a_r({\bf k}) + v_r b_r^{\dagger}(-{\bf k}) = \frac{1}{\sqrt{2}}
			\begin{pmatrix}
				\left[\left(A a_r - B^* b_r^{\dagger}\right)\hat{u}_+ + \left(Ba_r + A^* b_r^{\dagger}\right)\hat{v}_+\right]\psi_r \\%
				\left[\left(A a_r - B^* b_r^{\dagger}\right)\hat{u}_- + \left(Ba_r + A^* b_r^{\dagger}\right)\hat{v}_-\right]\psi_r
			\end{pmatrix}
	\end{equation}
	Comparing (\ref{equation:chi:pure:positive}) and (\ref{equation:bogoliubov:1}) shows that we need to define $\alpha = A$ and $\beta = -B^*$ in order for 
	\begin{equation}
		\label{equation:expansion:pure:freq:base}
		u_r a_r({\bf k}) + v_r b_r^{\dagger}(-{\bf k}) = \frac{1}{\sqrt{2}}
		\begin{pmatrix}
			\hat{u}_+\psi_r\\%
			\hat{u}_-\psi_r
		\end{pmatrix}
		\hat{a}_r({\bf k}) + \frac{1}{\sqrt{2}}
		\begin{pmatrix}
			\hat{v}_+\psi_r\\%
			\hat{v}_-\psi_r
		\end{pmatrix}
		\hat{b}_r^{\dagger}(-{\bf k}).
	\end{equation}
	Using this and the result of section \ref{subsection:wkb:bogo} we have the Bogoliubov coefficients $\beta_{\bf k} = - e^{-\frac{\pi\nu_0^2}{2\nu_1^2}}$ and $\alpha_{\bf k}$ a small phase rotation.
		
\section{Power spectrum and non-Gaussianity from fermion production}
\label{section:powerspectrum:nongaussianity}
In this section we compute the contributions to the inflaton correlation function from the fermion production events and compare them with those from the boson production events. As argued earlier, if the bosons and fermions are super partners and satisfy $|M_b|^2 = |M_f|^2 = M^2$, their radiative contributions from loop diagrams involving inflaton cancel with each other, therefore we only consider effects from particle production.

The general inflaton $N$-point function can be computed using the $in$-$in$ formalism 
\begin{align}
	\label{equation:N:point:original}
	\langle in | &\delta\phi_{{\bf k}_1}(0)\cdots \delta\phi_{{\bf k}_N} (0)| in \rangle = \nonumber \\%
	& \langle in | \left[\bar{T}\exp\left(i\int^0_{-\infty} d\eta H_I(\eta)\right)\right]~\delta\phi_{{\bf k}_1}(0) \cdots \delta\phi_{{\bf k}_N}(0) ~\left[T\exp\left(-i\int^0_{-\infty} d\eta H_I(\eta)\right)\right] | in \rangle,
\end{align}
where it is understood that the $\delta\phi$'s on the left hand side are evolved in the full theory and those on the right hand side are the free fields. $H_I(\eta)$ is the interaction Hamiltonian, which from (\ref{equation:interaction:lag:chi}) can be easily derived as 
\begin{align}
	\label{equation:interaction:hamiltonian}
	H_I(\eta) &= a(\eta)\frac{\delta M}{\delta\phi}\Big|_{\eta} \int d^3x \delta\phi(\eta) \bar{\chi}\chi(\eta) + \frac{1}{2} a(\eta)\frac{\delta^2 M}{\delta\phi^2}\Big|_{\eta}\int d^3x \delta\phi(\eta)^2 \bar{\chi}\chi(\eta) + \cdots \nonumber \\%
		&=  a(\eta)\frac{\delta M}{\delta\phi}\Big|_{\eta}\int_{\bf k} \delta\phi_{\bf k}(\eta)\bar{\chi}\chi_{-\bf k}(\eta) + \frac{1}{2} a(\eta)\frac{\delta^2 M}{\delta\phi^2}\Big|_{\eta}\int_{{\bf k}'_1, {\bf k}'_2}\delta\phi_{{\bf k}'_1}(\eta)\delta\phi_{{\bf k}'_2}(\eta)\bar{\chi}\chi_{-{\bf k}'_1 - {\bf k}'_2}(\eta) + \cdots.
\end{align}
The $|in\rangle$ state is the initial state which is vacuum in terms of the $a_r({\bf k})$ and $b_r({\bf k})$ operators. After the $n$th production event, it can be expressed as the squeezed state in the Fock space defined by the $\hat{a}_r({\bf k})$ and $\hat{b}_r({\bf k})$ operators
\begin{equation}
	\label{equation:squeezed:state:fermion}
	| in \rangle = \mathcal{N}\exp\left(\sum_{r = \pm 1}\int_{\bf k}\frac{\beta_{\bf k}}{\alpha^*_{\bf k}}\hat{a}_r^{\dagger}({\bf k})\hat{b}^{\dagger}_r(-{\bf k})\right) | \hat{0}\rangle 
		\simeq	\left(1 + \sum_r \int_{\bf k}\frac{\beta_{\bf k}}{\alpha^*_{\bf k}}\hat{a}_r^{\dagger}({\bf k})\hat{b}^{\dagger}_r(-{\bf k})\right)|\hat{0}\rangle,
\end{equation}
where $\mathcal{N}$ is a normalization constant and $|\hat{0}\rangle$ is the vacuum state in the Fock space defined by $\hat{a}_r({\bf k})$ and $\hat{b}_r({\bf k})$. The last approximation is due to the fact that $\beta$ is exponentially small and $\alpha$ is a small phase rotation. This can be checked by the following result
\begin{align}
	\langle \hat{a}^{\dagger}_{r_1}({\bf k}_1)\hat{a}_{r_2}({\bf k}_2) &\simeq 
			\langle\hat{0}| \left(1 + \sum_{s_1}\int_{{\bf q}_1}\frac{\beta^*_{{\bf q}_1}}{\alpha_{{\bf q}_1}}\hat{b}_{s_1}(-{\bf q}_1)\hat{a}_{s_1}({\bf q}_1)\right)\hat{a}^{\dagger}_{r_1}({\bf k}_1)\hat{a}_{r_2}({\bf k}_2) \nonumber \\%
				&\times \left(1 + \sum_{s_2}\int_{{\bf q}_2}\frac{\beta_{{\bf q}_2}}{\alpha^*_{{\bf q}_2}}\hat{b}_{s_2}(-{\bf q}_2)\hat{a}_{s_2}({\bf q}_2)\right)|\hat{0}\rangle \nonumber \\%
				& = \sum_{s_1, s_2}\int_{{\bf q}_1, {\bf q}_2}\frac{\beta^*_{{\bf q}_1}\beta_{{\bf q}_2}}{\alpha_{{\bf q}_1}\alpha^*_{{\bf q}_2}}\langle|\hat{0}| \hat{b}_{s_1}(-{\bf q}_1)\hat{a}_{s_1}({\bf q}_1)\hat{a}^{\dagger}_{r_1}({\bf k}_1)\hat{a}_{r_2}({\bf k}_2)\hat{a}^{\dagger}_{s_2}({\bf q}_2)\hat{b}^{\dagger}_{s_2}(-{\bf q}_2)|\hat{0}\rangle \nonumber \\%
				& \simeq \delta_{r_1r_2}\delta_{{\bf k}_1{\bf k}_2}|\beta_{k_1}|^2,
\end{align}
which gives the usual particle density $|\beta_{k}^2|$.

To compute the inflaton correlation functions we would also need $\bar{\chi}\chi_{\bf k}(\eta)$. Using the expansion (\ref{equation:chi:expansion}) and (\ref{equation:expansion:pure:freq:base}) we have 
\begin{align}
	\bar{\chi}\chi_{\bf k}(\eta) = \int_{{\bf k}'}&\sum_{r, s} \Big[
		\hat{u}^{\dagger}_r({\bf k}')	\gamma^0 \hat{u}_s({\bf k} + {\bf k}') \hat{a}^{\dagger}_r({\bf k}')\hat{a}_s({\bf k} + {\bf k}') 
		+ \hat{v}^{\dagger}_r({\bf k}')\gamma^0\hat{v}_s({\bf k} + {\bf k}')\hat{b}_r(-{\bf k}')\hat{b}^{\dagger}_s(-{\bf k}-{\bf k}')\nonumber \\%
		&+ \hat{u}^{\dagger}_r({\bf k}')\gamma^0\hat{v}_s({\bf k} + {\bf k}')\hat{a}^{\dagger}_r({\bf k}')\hat{b}^{\dagger}_s(-{\bf k}-{\bf k}') 
		+ \hat{v}^{\dagger}_r({\bf k}')\gamma^0\hat{u}_s({\bf k}+{\bf k}')\hat{b}_r(-{\bf k}')\hat{a}_s({\bf k} + {\bf k}')
	\Big],
\end{align}
where $\hat{u}_r({\bf k})$ and $\hat{v}_r({\bf k})$ are defined as 
\begin{equation}
	\hat{u}_r = \frac{1}{\sqrt{2}}
		\begin{pmatrix}
			\hat{u}_+\psi_r \\%
			\hat{u}_- \psi_r
		\end{pmatrix}
		\qquad
	\hat{v}_r = \frac{1}{\sqrt{2}}
		\begin{pmatrix}
			\hat{v}_+\psi_r \\%
			\hat{v}_-\psi_r
		\end{pmatrix}
\end{equation}
and $\hat{u}_{\pm}, \hat{v}_{\pm}$ are defined in (\ref{equation:definition:u:pm}). As we will show in the following, ${\bf k}'$ integrals contain $\mathcal{O}(|\beta|^2)$ or $\mathcal{O}(|\beta|)$ factors and peak at order $a_n\sqrt{g|\dot{\phi}|}$. Since the external momenta ${\bf k}$ are of order $H$, which is parametrically smaller than $a_n\sqrt{g|\dot{\phi}|}$, we can approximate ${\bf k}'\pm {\bf k}$ by ${\bf k}'$. Therefore we have 
\begin{align}
	\hat{u}^{\dagger}_r({\bf k}')\gamma^0 \hat{u}_s({\bf k} + {\bf k}') 
			&=\frac{1}{2}\left[\hat{u}^*_{+}({\bf k}')\hat{u}_+({\bf k} + {\bf k}')\psi^{\dagger}_r({\bf k}')\psi_s({\bf k}+{\bf k}') - \hat{u}^*_{-}({\bf k}')\hat{u}_{-}({\bf k} + {\bf k}')\psi^{\dagger}_r({\bf k}')\psi_s({\bf k}+{\bf k}') \right] \nonumber \\%
			&\simeq \frac{1}{2}\left[\left(1 + \frac{aM}{\omega}\right)\delta_{rs} - \left(1-\frac{aM}{\omega}\right)\delta_{rs}\right] = \frac{aM}{\omega}\delta_{rs} \simeq \delta_{rs},
\end{align}
where the last approximation is due to the fact that the ${\bf k}'$ integral peak location $a_n\sqrt{g|\dot{\phi}|}$ is parametrically smaller than $aM$. 
Similarly
\begin{align}
	&\hat{v}^{\dagger}_r({\bf k}')\gamma^0\hat{v}_s({\bf k}+{\bf k}') \simeq \delta_{rs}, \\%
	&\hat{u}^{\dagger}_r({\bf k}')\gamma^0\hat{v}_s({\bf k}+{\bf k}') \simeq \frac{k'}{aM}e^{2i\int\omega}\delta_{rs},\\%
	&\hat{v}^{\dagger}_r({\bf k}')\gamma^0\hat{u}_s({\bf k}+{\bf k}') \simeq \frac{k'}{aM}e^{-2i\int\omega}\delta_{rs},
\end{align}
where again $\int \omega$ is a short hand notation for $\int_{\eta_n}^{\eta}d\eta' \omega(\eta', {\bf k}')$.
Combining these results we have 
\begin{align}
	\label{equation:source:fermion}
	\bar{\chi}\chi_{\bf k}(\eta) &\simeq \int_{{\bf k}'}\sum_r
		\Big[\hat{a}^{\dagger}_r({\bf k}')\hat{a}_r({\bf k}+{\bf k}') + \hat{b}^{\dagger}_r(-{\bf k}-{\bf k}')\hat{b}_r(-{\bf k}') \nonumber \\%
		&+\frac{k'}{aM}e^{2i\int\omega}\hat{a}^{\dagger}_r({\bf k}')\hat{b}^{\dagger}_r(-{\bf k}-{\bf k}') + \frac{k'}{aM}e^{-2i\int\omega}\hat{b}_r(-{\bf k}')\hat{a}_r({\bf k}+{\bf k}')
	\Big].
\end{align}

In order to make direct comparisons between the fermion and boson contributions we also compute the relevant quantities for the boson case. The interaction Hamiltonian is
\begin{equation}
	H_I(\eta) = \frac{1}{2}a^4(\eta)\frac{\delta M^2}{\delta\phi}\Big|_{\eta}\int_{\bf k}\delta\phi_{\bf k}(\eta)\chi^2_{-{\bf k}}(\eta) + \frac{1}{4}a^{4}(\eta)\frac{\delta^2 M^2}{\delta\phi^2}\Big|_{\eta}\int_{{\bf k}'_1,{\bf k}'_2}\delta\phi_{{\bf k}'_1}\delta\phi_{{\bf k}'_2}\chi^2_{-{\bf k}'_1-{\bf k}'_2}(\eta) + \cdots
\end{equation}
and the source field is
\begin{equation}
	\label{equation:source:boson}
	\chi^2_{\bf k}(\eta) \simeq \frac{a^{-3}(\eta)}{2M(\eta)}\int_{{\bf k}'}\left[2a^{\dagger}_{-{\bf k}'}a_{{\bf k}-{\bf k}'} + e^{-2i\int\omega}a_{{\bf k}'}a_{{\bf k}-{\bf k}'} + e^{2i\int\omega}a^{\dagger}_{-{\bf k}'}a^{\dagger}_{-{\bf k}+{\bf k}'}\right],
\end{equation}
where we have dropped the hats on the $a$ operators and from this point on it is understood that they mean the operators defined with respect to purely positive and negative frequency modes.

It is argued in \cite{Flauger:2016idt} that the oscillating terms in (\ref{equation:source:boson}) only contribute to $N$-point functions of $\delta\phi$ through $(N+2)$-point vertex insertions. We elaborate on this point in section \ref{section:interaction:vertices:profile}. Order $|\beta|^2$ contributions and order $|\beta|$ contributions are computed in section \ref{section:order:betasquare:contrib} and \ref{section:order:beta:contrib}, respectively.

\subsection{Contributions from various interaction vertex profiles}
\label{section:interaction:vertices:profile}
The oscillating terms in (\ref{equation:source:fermion}) and (\ref{equation:source:boson}) have either two creation or two annihilation operators and can contract with the two annihilation or creation operators from the squeezed state. Therefore it is possible to have only one $\beta$ factor when evaluating the $\delta\phi$ correlators as opposed to the order $|\beta|^2$ contributions when only using the $a^{\dagger}a$ terms. It is argued in \cite{Flauger:2016idt} that such order $\beta$ contribution only shows up when using $(N+2)$ point vertex for  computing $N$ point function. The reason is that all the other use of such terms do not resonate and therefore their contribution is negligible. Since such argument applies to both boson and fermion, we will work with boson in this subsection.

A generic term after expanding the $N$ point function (\ref{equation:N:point:original}) is
\begin{align}
	\label{equation:N:point:generic:term}
	& i^{l_1}(-i)^{l_2}\int^0_{-\infty}d\eta_{l_1}\int^{\eta_{l_1}}_{-\infty}d\eta_{l_1 - 1}\cdots \int^{\eta_2}_{-\infty}d\eta_1\int^{0}_{-\infty}d\zeta_{l_2}\cdots \int^{\zeta_2}_{-\infty}d\zeta_1 \nonumber \\%
	& \left(\frac{1}{2 o_{l_1}!}\frac{\delta^{o_{l_1}}M^2}{\delta\phi^{o_{l_1}}}\Big|_{\eta_{l_1}}a^4(\eta_{l_1})\cdots \frac{1}{2o_1!}\frac{\delta^{o_1}M^2}{\delta\phi^{o_1}}\Big|_{\eta_1}a^4(\eta_1)\right)
		\left(\frac{1}{2o'_{l_2}!}\frac{\delta^{o'_{l_2}}M^2}{\delta\phi^{o'_{l_2}}}\Big|_{\zeta_{l_2}}\cdots \frac{1}{2o'_1!}\frac{\delta^{o'_1}M^2}{\delta\phi^{o'_1}}\Big|_{\zeta_1}a^4(\zeta_1)\right) \nonumber \\%
	& \langle\delta\phi_{{\bf k}'_{1,1}}(\eta_1)\cdots\delta\phi_{{\bf k}'_{1, o_1}}(\eta_1)\cdots \delta\phi_{{\bf k}'_{l_1, o_{l_1}}}(\eta_{l_1})\delta\phi_{{\bf k}_1}(0)\cdots \delta\phi_{{\bf k}_N}(0) \delta\phi_{{\bf p}_{l_2, o_{l_2}}}(\zeta_{l_2})\cdots \delta\phi_{{\bf p}_{1, o'_1}}(\zeta_1)\cdots \delta\phi_{{\bf p}_{1, 1}}(\zeta_1)\rangle \nonumber \\%
	&\langle \chi^2_{-\sum_{i=1}^{o_1}{\bf k}'_{1,i}}(\eta_1)\cdots \chi^2_{-\sum_{i=1}^{o_{l_1}}{\bf k}'_{l_1,i}}(\eta_{l_1})\chi^2_{-\sum_{i=1}^{o'_{l_2}}{\bf p}_{l_2,i}}(\zeta_{l_2})\cdots \chi^2_{-\sum_{i=1}^{o'_1}{\bf p}_{1,i}}(\zeta_1)\rangle,
\end{align}
where there are $l_1$ insertions from the anti-time-ordered operator and $l_2$ insertions from the time-ordered operator, with the $i$th vertices being $(o_i + 2)$ and $(o'_i + 2)$ point vertices, respectively. To contract each of the $\delta\phi_{\bf k}(0)$ operators with a $\delta\phi_{\bf k}(\eta)$ or $\delta\phi_{\bf k}(\zeta)$ operator from $H_I$ we also have $\sum_{i=1}^{l_1}o_i + \sum_{i=1}^{l_2}o'_i = N$. Also notice that since on the right hand side of (\ref{equation:N:point:original}), $\delta\phi_{\bf k}$'s are evolved using the free inflaton Hamiltonian the expectation value factorizes as in (\ref{equation:N:point:generic:term}) and the $\delta\phi_{\bf k}$ correlator is computed in the $\delta\phi$ vacuum state while the $\chi^2_{\bf k}$ correlator is computed in the $\chi$ squeezed state.

In order to look for the contributions of order $\beta$ or even $\mathcal{O}(1)$, we first consider the case where the first insertion comes from the anti-time-ordered operator, i.e., $l_1 > 0$. Similar to (\ref{equation:squeezed:state:fermion}) we have the squeezed state for the boson case
\begin{equation}
	\label{equation:squeezed:state:boson}
	|in \rangle = \mathcal{N}\exp\left(\int_{\bf k}\frac{\beta_{\bf k}}{2\alpha^*_{\bf k}}a^{\dagger}_{\bf k}a^{\dagger}_{-{\bf k}}\right)|0\rangle 
		\simeq \left(1 + \int_{\bf k}\frac{\beta_{\bf k}}{2\alpha^*_{\bf k}}a^{\dagger}_{\bf k}a^{\dagger}_{-{\bf k}}\right)|0\rangle.
\end{equation}
If there is no $\frac{\beta^*}{\alpha}aa$ on the left of the $\chi^2$ operators in (\ref{equation:N:point:generic:term}), then the first $\chi^2$ factor $\chi^2_{-\sum_{i=1}^{o_1}{\bf k}'_{1, i}}$ must contribute an $aa e^{-2i\mu (t_1-t_p)}$ factor\footnote{$t_p$ denotes the time of minimum $\chi$ mass.}. Since $\delta\phi_{{\bf k}'_{1,i}}(\eta_1)$ for $\forall i$ have to contract with some $\delta\phi_{{\bf k}}(0)$ later in the product, they all contribute an oscillating factor $e^{-ik'_{1, i}\eta_1}$. Using $t_1-t_p = -\frac{1}{H}\log\frac{\eta_1}{\eta_p}$, we have the oscillating part of the $\eta_1$ integral as $\exp\left(-i\sum_{i}^{o_1}k'_{1,i}\eta_1 + i\frac{2\mu \pm \omega}{H}\log\frac{\eta_1}{\eta_p}\right)$ where $e^{\pm i\frac{\omega}{H}}$ comes from the mass derivatives. It has no resonance in the parameter region we work with where $\mu > \omega$. Therefore there must be a $\frac{\beta^*}{\alpha}aa$ factor on the left. In order to get contribution of order $\mathcal{O}(\beta)$ we cannot have a $\frac{\beta}{\alpha^*}a^{\dagger}a^{\dagger}$ on the right and the last $\chi^2$ insertion must give an $a^{\dagger}a^{\dagger}e^{2i\mu(t-t_p)}$ factor.

If the last vertex is from the time-ordered operator ($l_2 > 0$), then $\delta\phi_{{\bf p}_{1, i}}(\zeta_1)$ must contribute an $e^{i\sum_{i=1}^{o'_1} p_{1, i}\zeta_1}$ factor. It does not resonate with the $\exp\left(-i\frac{2\mu}{H}\log\frac{\zeta_1}{\eta_p}\right)$ factor from $\chi^2_{-\sum_i {\bf p}_{1, i}}(\zeta_1)$ for the same reason as in the previous paragraph. If the last vertex is from the anti-time-ordered operator ($l_2 = 0$), then the oscillating part of its time integral is
\begin{equation}
	\sin\left(\frac{\omega}{H}\log\frac{\eta_{l_1}}{\eta_p}\right)e^{- i \sum_{i=1}^{o_{l_1}} k'_{l_1, i}\eta_{l_1}-i\frac{2\mu}{H}\log\frac{\eta_{l_1}}{\eta_p}},
\end{equation}
which resonates at $-(\sum_{i}k'_{l_1, i})\eta_{l_1} = \frac{2\mu\pm \omega}{H}$. All the other time integrals resonate at either $k\eta = -\frac{\omega}{H} > -\frac{2\mu\pm \omega}{H}$ or $-\frac{2\mu\pm\omega}{H}$, which means the resonance point of $\eta_{l_1}$ is the earliest. However the anti-time-ordering requires $\eta_{l_1}$ to be the latest, so there is still no resonance in the integral region.

Next we consider the case where the first insertion is from the time-ordered operator (i.e., $l_1=0$). Similar to the previous case there must be a $\frac{\beta}{\alpha^*}a^{\dagger}a^{\dagger}$ factor on the right and no $\frac{\beta^*}{\alpha}aa$ on the left. Therefore the first insertion must give $aa\exp\left(i\frac{2\mu}{H}\log\frac{\zeta_{l_2}}{\eta_p}\right)$ and its time integral resonates at $k\zeta_{l_2} = -\frac{2\mu\pm\omega}{H}$. Again due to the time-ordering requirement $\zeta_{l_2} > \zeta_{l_2 - 1} > \cdots > \zeta_1$ there is no resonance in the integral region.

The analysis above shows that the only way for the $\chi^2$'s to contribute order $\beta$ oscillating factor is to have only one $\chi^2$, i.e., bring down only one $H_{N+2}$ vertex. In the next subsection we compute the contributions to the $N$ point functions from $N$ three point vertices and show that the fermion case and boson case are the same except for a numerical factor coming from the helicity and anti-particle doubling. Since for all the other ways of using various $H_I$ terms (other than using only one $H_{N+2}$), only $a^{\dagger}a$ and $b^{\dagger}b$ contribute, it is clear that contributions to the $N$ point functions should be the same between the fermion case and the boson case in those situations. After that we compute the contributions from only using one $H_{N+2}$ vertex and show that the order $\beta$ contribution in the fermion case is further suppressed than in the boson case.

\subsection{Order $|\beta|^2$ contributions}
\label{section:order:betasquare:contrib}
The contributions to the inflaton $N$ point function from bringing down $N$ three point vertices can be organized as \cite{Weinberg:2005vy}
\begin{equation}
	\label{equation:order:betasquare:orig}
	i^N\int^0_{-\infty}d\eta_{N}\cdots \int^{\eta_2}_{-\infty}d\eta_1\langle\left[H_{3pt}(\eta_1), \cdots, \left[H_{3pt}(\eta_{\eta_N}), \delta\phi_{{\bf k}_1}(0)\cdots\delta\phi_{{\bf k}_N}(0)\right]\cdots\right]\rangle,
\end{equation}
where 
\begin{equation}
	H_{3pt}(\eta) = a(\eta)\frac{\delta M}{\delta\phi}\Big|_{\eta}\int_{{\bf k}'}\delta\phi_{{\bf k}'}(\eta)\bar{\chi}\chi_{-{\bf k}'}(\eta).
\end{equation}
The result in subsection \ref{section:interaction:vertices:profile} shows that only classical source terms $a^{\dagger}a$ and $b^{\dagger}b$ in (\ref{equation:source:fermion}) have contribution in this case and we will see in the following that expectation values of the product of those operators do not rely on their order. Therefore the expectation value of (\ref{equation:order:betasquare:orig}) factorizes as 
\begin{align}
	\label{equation:order;betasquare:fermion}
	i^N &\int_{-\infty}^0 d\eta_N\cdots\int_{-\infty}^{\eta_2}d\eta_1\left(\prod_{i=1}^N a(\eta_i)\frac{\delta M}{\delta\phi}\Big|_{\eta_i}\right)\int_{{\bf k}'_1, \cdots, {\bf k}'_N}\langle\bar{\chi}\chi_{-{\bf k}'_1}(\eta_1)\cdots \bar{\chi}\chi_{-{\bf k}'_N}(\eta_N)\rangle_{\mathrm{cl}} \nonumber \\%
		&\times \langle[\delta\phi_{{\bf k}'_1}(\eta_1), \cdots, [\delta\phi_{{\bf k}'_N}(\eta_N), \delta\phi_{{\bf k}_1}(0)\cdots \delta\phi_{{\bf k}_N}(0)]\cdots]\rangle,
\end{align}
where the subscript ``cl'' means computing expectation value only using the classical source terms. The expectation value of the $\delta\phi$ commutator can be proven to be 
\begin{align}
	\label{equation:N:deltaphi:commutator}
	&[\delta\phi_{{\bf k}'_1}(\eta_1), \delta\phi_{{\bf k}_1}(0)]~[\delta\phi_{{\bf k}'_2}(\eta_2), \delta\phi_{{\bf k}_2}(0)]\cdots[\delta\phi_{{\bf k}'_N}(\eta_N), \delta\phi_{{\bf k}_N}(0)] \nonumber \\%
	&+ (\mathrm{all~other~permutations~of~}\delta\phi_{{\bf k}_1}(0), \cdots,  \delta\phi_{{\bf k}_N}(0))
\end{align}
by mathematical induction. First note that from 
\begin{equation}
	\label{equation:inflation:mode:expansion}
	\delta\phi_{\bf k}(\eta) = a_{\bf k}u_k(\eta) + a^{\dagger}_{-\bf k}u^*_{k}(\eta), \quad 
		u_k(\eta) = \frac{H}{\sqrt{2k^3}}(i-k\eta) e^{-ik\eta}
\end{equation}
we have 
\begin{equation}
	[\delta\phi_{{\bf k}'}(\eta), \delta\phi_{\bf k}(0)] = \delta_{-{\bf k}',{\bf k}}\frac{-iH^2}{k^3}[\sin(k\eta) - k\eta\cos(k\eta)] \equiv \delta_{-{\bf k}',{\bf k}}iG_{k}(0, \eta)
\end{equation}
and it is a $c$-number. Then in the case of $N=2$ it is easy to show directly that 
\begin{align}
	\langle\left[\delta\phi_{{\bf k}'_1}(\eta_1), \left[\delta\phi_{{\bf k}'_2}(\eta_2),  \delta\phi_{{\bf k}_1}(0)\delta\phi_{{\bf k}_2}(0)\right]~\right]\rangle &= [\delta\phi_{{\bf k}'_1}(\eta_1), \delta\phi_{{\bf k}_1}(0)]~[\delta\phi_{{\bf k}'_2}(\eta_2), \delta\phi_{{\bf k}_2}(0)] \nonumber \\%
			&+ [\delta\phi_{{\bf k}'_1}(\eta_1), \delta\phi_{{\bf k}_2}(0)]~[\delta\phi_{{\bf k}'_2}(\eta_2), \delta\phi_{{\bf k}_1}(0)].
\end{align}
Next we assume that (\ref{equation:N:deltaphi:commutator}) holds for $N-1$. By using the identity
\begin{align}
	[O, P_1\cdots P_N]  = &\Big(P_1\cdots P_{N-1}[O, P_N]\Big) + \cdots + \Big( P_1\cdots P_{i-1}[O, P_i]P_{i+1}\cdots P_N \Big) \nonumber \\%
		& + \cdots + \Big( [O, P_1]P_2 \cdots P_N\Big)
\end{align}
for any operator $O$ and $P_i$ we have
\begin{align}
	& \Big\langle\left[\delta\phi_{{\bf k}'_1}(\eta_1), \cdots, \left[\delta\phi_{{\bf k}'_N}(\eta_N), \delta\phi_{{\bf k}_1}(0)\cdots \delta\phi_{{\bf k}_N}(0)\right]\cdots\right]\Big\rangle \nonumber \\%
	= &\Big\langle\Big[\delta\phi_{{\bf k}'_1}(\eta_1), \cdots, \Big[\delta\phi_{{\bf k}'_{N-1}}(\eta_{N-1}), ~\delta\phi_{{\bf k}_1}(0)\cdots\delta\phi_{{\bf k}_{N-1}}(0)~\left[\delta\phi_{{\bf k}'_N}(\eta_N), \delta\phi_{{\bf k}_N}(0)\right] + \cdots \nonumber \\%
		& \qquad+ \left[\delta\phi_{{\bf k}'_N}(\eta_N), \delta\phi_{{\bf k}_1}(0)\right]~\delta\phi_{{\bf k}_2}(0)\cdots\delta\phi_{{\bf k}_N}(0)\Big]\cdots\Big]\Big\rangle \nonumber \\%
	 = &\left[\delta\phi_{{\bf k}'_N}(\eta_N), \delta\phi_{{\bf k}_N}(0)\right]~\Big\langle\left[\delta\phi_{{\bf k}'_1}(\eta_1), \cdots,~[\delta\phi_{{\bf k}'_{N-1}}(\eta_{N-1}), \delta\phi_{{\bf k}_1}(0)\cdots \delta\phi_{{\bf k}_{N-1}}(0)]~\cdots\right]\Big\rangle + \cdots \nonumber \\%
	& \qquad +\left[\delta\phi_{{\bf k}'_N}(\eta_N), \delta\phi_{{\bf k}_1}(0)\right]~\Big\langle\left[\delta\phi_{{\bf k}'_1}(\eta_1), \cdots,~[\delta\phi_{{\bf k}'_{N-1}}(\eta_{N-1}), \delta\phi_{{\bf k}_2}(0)\cdots\delta\phi_{{\bf k}_N}(0)]\cdots \right]\Big\rangle \nonumber \\%
	 = &\left[\delta\phi_{{\bf k}'_1}(\eta_1), \delta\phi_{{\bf k}_1}(0)\right]~\left[\delta\phi_{{\bf k}'_2}(\eta_2), \delta\phi_{{\bf k}_2}(0)\right]\cdots\left[\delta\phi_{{\bf k}'_N}(\eta_N), \delta\phi_{{\bf k}_N}(0)\right] \nonumber \\%
	& \qquad + (\mathrm{all~other~permutations~of~}\delta\phi_{{\bf k}_1}(0), \cdots,  \delta\phi_{{\bf k}_N}(0)),
\end{align}
where in the last equality we used the induction assumption that (\ref{equation:N:deltaphi:commutator}) holds up to $N-1$. This completes the proof. With this result (\ref{equation:order;betasquare:fermion}) can be written as 
\begin{align}
	(-1)^N &\int^0_{-\infty}\cdots\int^{\eta_2}_{-\infty}d\eta_1\left(\prod_{i=1}^N a(\eta_i)\frac{\delta M}{\delta\phi}\Big|_{\eta_i}\right) \nonumber \\%
	& \times\sum_{\pi}G_{{\bf k}_1}(0, \eta_{\pi_1})\cdots G_{{\bf k}_N}(0, \eta_{\pi_N})\langle\bar{\chi}\chi_{{\bf k}_1}(\eta_{\pi_1})\cdots \bar{\chi}\chi_{{\bf k}_N}(\eta_{\pi_N})\rangle_{\mathrm{cl}},
\end{align}
where the summation $\pi$ is over all permutations of $1, 2, \cdots, N$. Each of the permutation can be renamed such that the integrand becomes $G_{{\bf k}_1}(0, \eta_1)\cdots G_{{\bf k}_N}(0, \eta_N)\langle\bar{\chi}\chi_{{\bf k}_1}(\eta_1)\cdots \bar{\chi}\chi_{{\bf k}_N}(\eta_N)\rangle_{\mathrm{cl}}$ and as a result it has a different time ordering from the one in the original integral. Summing over all the permutations covers the entire space spanned by $\eta_1, \cdots, \eta_N$ and therefore the integral simplifies as
\begin{equation}
	(-1)^N\int^0_{-\infty} d\eta_N \cdots \int^0_{-\infty}d\eta_1 \left(\prod_{i=1}^N a(\eta_i)\frac{\delta M}{\delta\phi}\Big|_{\eta_i}G_{{\bf k}_k}(0, \eta_i)\right)\langle\bar{\chi}\chi_{{\bf k}_1}(\eta_1)\cdots \bar{\chi}\chi_{{\bf k}_N}(\eta_N)\rangle_{\mathrm{cl}}.
\end{equation}

Next we need to compute the $\chi$ correlator
\begin{align}
	\langle\bar{\chi}\chi_{{\bf k}_1}(\eta_1)\cdots \bar{\chi}\chi_{{\bf k}_N}(\eta_N)\rangle_{\mathrm{cl}} &\simeq \int_{{\bf k}'_1, \cdots {\bf k}'_N}\sum_{r_1,\cdots, r_N}
		\langle \left[a^{\dagger}_{r_1}({\bf k}'_1)a_{r_1}({\bf k}_1 +  {\bf k}'_1) + b^{\dagger}_{r_1}(-{\bf k}_1-{\bf k}'_1)b_{r_1}(-{\bf k}'_1)\right] \nonumber \\%
			& \times \cdots \times \left[a^{\dagger}_{r_N}({\bf k}'_N)a_{r_N}({\bf k}_N + {\bf k}'_N) + b^{\dagger}_{r_N}(-{\bf k}_N-{\bf k}'_N)b_{r_N}(-{\bf k}'_N)\right]\rangle.
\end{align}
It is clear that whether each factor contributes an $a^{\dagger}a$ or $b^{\dagger}b$, the expectation value is always
\begin{align}
	& \langle c^{\dagger}_{r_1}({\bf k}'_1)c_{r_1}({\bf k}_1+ {\bf k}'_1) \cdots c^{\dagger}_{r_N}({\bf k}'_N)c_{r_N}({\bf k}_N+ {\bf k}'_N)\rangle \nonumber \\%
	\simeq & \sum_{s_1, s_2}\int_{{\bf q}_1, {\bf q}_2}\frac{\beta^*_{q_1}\beta_{q_2}}{\alpha_{q_1}\alpha_{q_2}} 
	\Big\langle b_{s_1}(-{\bf q}_1)a_{s-1}({\bf q}_1) c^{\dagger}_{r_1}({\bf k}'_1)c_{r_1}({\bf k}_1+ {\bf k}'_1) \nonumber \\%
			& \qquad \qquad \cdots c^{\dagger}_{r_N}({\bf k}'_N)c_{r_N}({\bf k}_N+ {\bf k}'_N)a^{\dagger}_{s_2}({\bf q}_2)b^{\dagger}_{s_2}(-{\bf q}_2)\Big\rangle \nonumber \\%
	\simeq & |\beta_{k'_1}|^2\delta_{r_1 r_2}\cdots \delta_{r_{N-1}r_N}\delta_{{\bf k}'_1{\bf k}'_2}\cdots \delta_{{\bf k}'_{N-1}, {\bf k}'_{N}}\delta({\bf k}_T),
\end{align}
where $c$ can be either $a$ or $b$ and ${\bf k}_T = \sum_{i=1}^N {\bf k}_i$. Combining the $2^N$ of such terms we have 
\begin{equation}
	\langle\bar{\chi}\chi_{{\bf k}_1}(\eta_1)\cdots \bar{\chi}\chi_{{\bf k}_N}(\eta_N)\rangle_{\mathrm{cl}} \simeq 2^{N+1} \left(\int_{{\bf k}'_1}|\beta_{k'_1}|^2\right)\delta({\bf k}_T) = 2^{N+1} a_n^3\bar{n}_{\chi}\delta{{\bf k}_T},
\end{equation}
where we have used 
\begin{equation}
	\label{equation:n:def}
	\int_{{\bf k}'_1}|\beta_{k'_1}|^2 = a^3_n \int\frac{d^3p}{(2\pi)^3}\exp\left(-\frac{\pi(p^2 + \tilde{\mu^2})}{g|\dot{\phi}|}\right) = a^3_n \frac{(g|\dot{\phi}|)^{3/2}}{(2\pi)^3}e^{-\frac{\pi\tilde{\mu}^2}{g\dot{\phi}}} \equiv a_n^3 \bar{n}_{\chi}.
\end{equation}
Also note the extra factor of 2 that comes from the helicity summation, which the boson case does not have. Define
\begin{equation}
	h(k\eta_n) = \int^0_{\eta_n}\frac{d\eta}{\eta}\left(\sin(k\eta) - k\eta\cos(k\eta)\right)\frac{\delta M}{\delta\phi}\Big|_{\eta}
\end{equation}
then summing over different production events gives the contributions from the fermion three point vertices
\begin{equation}
	\langle\delta\phi_{{\bf k}_1}(0)\cdots\delta\phi_{{\bf k}_N}(0)\rangle \supset (-2)^{N+1}\delta({\bf k}_T)\frac{\bar{n}_{\chi}}{H^3}H^{N+3}\sum_n(H\eta_n)^{-3}\prod_{i=1}^N\frac{h(k_i\eta_n)}{k_i^3}.
\end{equation}
To be more concrete, in the rest of this paper we will consider a situation similar to that of \cite{Flauger:2016idt} with an approximate discrete shift symmetry, with production events evenly spaced in proper time $t$, corresponding to conformal times
\begin{equation}
	\eta_n = -\frac{1}{H}\exp\left[\frac{2\pi H}{w}\left(n + \frac{\gamma}{2\pi}\right)\right],
\end{equation}
with a constant $\gamma$ and frequency $w = |\dot{\phi}|/f$ derived from (\ref{equation:mass:sqaure:cosine}.

Using the boson three point vertex
\begin{equation}
	H_{3pt}(\eta) = \frac{1}{2}a^4(\eta)\frac{\delta M^2}{\delta\phi}\Big|_{\eta}\int_{{\bf k}'}\delta\phi_{{\bf k}'}(\eta)\chi^2_{-{\bf k}'}(\eta)
\end{equation}
and (\ref{equation:source:boson}) we can get the same contributions in the boson case
\begin{equation}
	\langle\delta\phi_{{\bf k}_1}(0)\cdots\delta\phi_{{\bf k}_N}(0)\rangle \supset (-2)^{N-1}\delta({\bf k}_T)\frac{\bar{n}_{\chi}}{H^3}H^{N+3}\sum_n(H\eta_n)^{-3}\prod_{i=1}^N\frac{h(k_i\eta_n)}{k_i^3}.
\end{equation}
except that it is smaller than the fermion case by a factor of $4$, which comes from the helicity doubling and anti-particle doubling. 

In the next subsection we will show that 
\begin{equation}
	\frac{\delta^N M}{\delta\phi^N}\simeq \frac{1}{2M}\frac{\delta^N M^2}{\delta\phi^N}.
\end{equation}
From this fact and the discussion above one can expect that the contributions from interaction vertex configurations including higher point vertices would also be the same for the fermion case and boson case (except for the case where only one $H_{N+2}$ is brought down for $N$ point function as shown in section \ref{section:order:beta:contrib}). This is because in both cases the $\delta\phi$ correlators and $\chi^2$ (or $\bar{\chi}\chi$) correlators give the same result and they are combined in the same way. While it is worth exploring the possible contributions from various insertion profiles using higher point vertices, we note that if one is working in the same parameter regime as outlined in \cite{Flauger:2016idt} the contributions from $(N+2)$ three point fermion vertices to $N$ point functions will dominate over other contributions involving higher point vertices. 

\subsection{Order $\beta$ contributions}
\label{section:order:beta:contrib}
In this subsection we consider the contributions from only bringing down one $H_{N+2}$ vertex for $N$ point functions. As discussed in section \ref{section:interaction:vertices:profile} such contributions contain parts that are of order $\beta$. For both the fermion case and the boson case they are
\begin{equation}
	\label{equation:order:beta:orig}
	i\int_{-\infty}^0 d\eta_1\langle\left[H_{N+2}(\eta_1), \delta\phi_{{\bf k}_1}(0)\cdots\delta\phi_{{\bf k}_N}(0)\right]\rangle.
\end{equation}
In the fermion case, the $(N+2)$ point vertices are
\begin{equation}
	H_{N+2}(\eta) = a(\eta)\frac{1}{N!}\frac{\delta^N M}{\delta\phi^N}\Big|_{\eta}\int_{{\bf k}'_1, \cdots, {\bf k}'_N}\delta\phi_{{\bf k}'_1}(\eta)\cdots \delta\phi_{{\bf k}'_N}(\eta)\bar{\chi}\chi_{-\sum_{i=1}^N{\bf k}'_i}(\eta),
\end{equation}
and (\ref{equation:order:beta:orig}) becomes
\begin{align}
	i & \int_{-\infty}^0 d\eta_1 a(\eta_1) \frac{1}{N!} \frac{\delta^N M}{\delta\phi^N}\Big|_{\eta_1} \int_{{\bf k}'_1, \cdots, {\bf k}'_N}
		\langle\bar{\chi}\chi_{-\sum_{i=1}^N{\bf k}'_i}(\eta_1)\rangle \nonumber \\%
		& \times \langle\left[\delta\phi_{{\bf k}'_1}(\eta_1)\cdots \delta\phi_{{\bf k}'_N}(\eta_1), \delta\phi_{{\bf k}_1}(0)\cdots\delta\phi_{{\bf k}_N}(0)\right]\rangle.
\end{align}
To compute the $\delta\phi$ commutator we first notice that in $\langle\delta\phi_{{\bf k}'_1}(\eta_1)\cdots \delta\phi_{{\bf k}'_N}(\eta_1)\delta\phi_{{\bf k}_1}(0)\cdots\delta\phi_{{\bf k}_N}(0)\rangle$ each of $\delta\phi_{{\bf k}'_i}(\eta_1)$ must contract with one $\delta\phi_{{\bf k}_j}(0)$ and therefore the former must be $a_{{\bf k}'_i}u_{k'_i}(\eta_1)$ and the later $a^{\dagger}_{-{\bf k}_j}u^*_{k_j}(0)$. So the correlator is evaluated to be
\begin{align}
	\langle\delta\phi_{{\bf k}'_1}(\eta_1)\cdots \delta\phi_{{\bf k}'_N}(\eta_1)\delta\phi_{{\bf k}_1}(0)\cdots\delta\phi_{{\bf k}_N}(0)\rangle = &\sum_{\pi}\delta_{-{\bf k}'_1{\bf k}_{\pi_1}}  \cdots  \delta_{-{\bf k}'_N{\bf k}_{\pi_N}}\nonumber \\%
		& \times u_{k_1}(\eta_1)u^*_{k_1}(0) \cdots u_{k_N}(\eta_1)u^*_{k_N}(0),
\end{align}
and the summation is again over all the possible permutation $\pi$ over $1, 2, \cdots, N$. And similarly
\begin{align}
	\langle\delta\phi_{{\bf k}_1}(0)\cdots\delta\phi_{{\bf k}_N}(0)\delta\phi_{{\bf k}'_1}(\eta_1)\cdots\delta\phi_{{\bf k}'_N}(\eta_1)\rangle = & \sum_{\pi}\delta_{-{\bf k}'_1{\bf k}_{\pi_1}}  \cdots  \delta_{-{\bf k}'_N{\bf k}_{\pi_N}}\nonumber \\%
	&\times u_{k_1}(0)u^*_{k_1}(\eta_1)\cdots u_{k_N}(0) u^*_{k_N}(\eta_1).
\end{align}
Plugging in 
\begin{equation}
	u_{k}(\eta)u^*_k(0) = \frac{H^2}{2k^3}\left(1 + ik\eta\right)e^{-ik\eta},\quad
	u_{k}(0) u^*_k(\eta) = \frac{H^2}{2k^3}\left(1 - ik\eta\right)e^{ik\eta},
\end{equation}
we have the fermion $(N+2)$ point vertex contribution
\begin{align}
	\label{equation:order:beta:fermion:interm}
	&\frac{iH^{2N}}{2^N\prod_{i=1}^N k_i^3}\int^0_{-\infty} d\eta_1 a(\eta_1)\frac{\delta^N M}{\delta\phi^N}\Big|_{\eta_1}\langle\bar{\chi}\chi_{{\bf k}_T}(\eta_1)\rangle \nonumber \\%
	&\left[\left(\prod_{i=1}^N (1+ik_i\eta_1)\right)e^{-ik_T\eta_1} -\left (\prod_{i=1}^N (1- ik_i\eta_1)\right)e^{ik_T\eta_1}\right],
\end{align}
where $k_T = \sum_{i=1}^N k_i$. Similarly the boson $(N+2)$ point vertex
\begin{equation}
	H_{N+2}(\eta) = \frac{1}{2}a^4(\eta)\frac{1}{N!}\frac{\delta^N M^2}{\delta\phi^N}\Big|_{\eta}\int_{{\bf k}'_1\cdots {\bf k}'_N}\delta\phi_{{\bf k}'_1}(\eta)\cdots\delta\phi_{{\bf k}'_N}(\eta)\chi^2_{-\sum_{i=1}^N {\bf k}'_i}(\eta)
\end{equation}
has the following contribution to the $N$ point correlator
\begin{align}
	\label{equation:order:beta:boson:interm}
	& \frac{iH^{2N}}{2^N\prod_{i=1}^N k_i^3}\int^0_{-\infty} d\eta_1 \frac{1}{2} a^4(\eta_1)\frac{\delta^N M^2}{\delta\phi^N}\Big|_{\eta_1}\langle\chi^2_{{\bf k}_T}(\eta_1)\rangle \nonumber \\%
	&\times \left[\left(\prod_{i=1}^N (1+ik_i\eta_1)\right)e^{-ik_T\eta_1} - \left(\prod_{i=1}^N (1- ik_i\eta_1)\right)e^{ik_T\eta_1}\right].
\end{align}
A comparison of (\ref{equation:order:beta:fermion:interm}) and (\ref{equation:order:beta:boson:interm}) shows the difference between the two cases could come from the difference in the two point functions of the produced particle fields. Indeed the $k'/aM$ factor in (\ref{equation:source:fermion}) and the fact that ${\bf k}'$ integral peaks around $a_n\sqrt{g|\dot{\phi}|} \ll aM$ indicate that the order $\beta$ contribution in the fermion case will be further suppressed by a factor of $\sqrt{g|\dot{\phi}|}/\mu$. We confirm this point in the following by evaluating (\ref{equation:order:beta:fermion:interm}) and (\ref{equation:order:beta:boson:interm}).

Let us start by computing $\langle\bar{\chi}\chi_{{\bf k}_T}(\eta_1)\rangle$. With $\langle a^{\dagger}_r({\bf k}')a_r({\bf k}_T + {\bf k}')\rangle = \langle	 b^{\dagger}_r(-{\bf k}_T - {\bf k}')b_r(-{\bf k}')\rangle \simeq |\beta_{k'}|^2\delta({\bf k}_T)$, $\langle a^{\dagger}_r({\bf k}')b^{\dagger}_r(-{\bf k}_T - {\bf k}')\rangle \simeq \beta^*_{k'}\delta({\bf k}_T)$ and $\langle b_r(-{\bf k}')a_r({\bf k}_T + {\bf k}') \simeq \beta_{k'}\delta({\bf k}_T)$ the mode expansion (\ref{equation:source:fermion}) becomes
\begin{equation}
	\langle\bar{\chi}\chi_{{\bf k}_T}(\eta_1) \rangle	=2\delta({\bf k}_T)\int_{{\bf k}'}\left[2|\beta_{k'}|^2 + \frac{k'}{\omega(k')}\left(e^{-2i\int\omega}\beta_{k'} + c.c. \right)\right].
\end{equation}
The integral over $\omega$ can be approximately evaluated as
\begin{equation}
	\int_{\eta_n}^{\eta}d\eta'\sqrt{k'^2 + a^2\left(\mu^2 + 2g^2f^2\cos\frac{\phi}{f}\right)} \simeq \int_{t_n}^{t} dt' \mu\left(1 + \frac{k'^2}{2\mu^2 a^2}\right) = \mu(t - t_n) + \frac{k'^2}{4a_n^2\mu H}A_n,
\end{equation}
where we have defined $A_n(\eta) = 1 - a_n^2/a(\eta)^2$. Using this and the definition (\ref{equation:n:def}) of particle density $\bar{n}_{\chi}$ the source expectation value is evaluated to be
\begin{align}
	\label{equation:source:fermion:expectation:value}
	\langle\bar{\chi}\chi_{{\bf k}_T}(\eta_1)\rangle  \simeq 4\delta({\bf k}_T)a_n^3 \bar{n}_{\chi}\left[1 - \frac{4}{\pi}\frac{a_n(g|\dot{\phi}|)^{1/2}}{a\mu}e^{\frac{\pi\tilde{\mu}^2}{2g|\dot{\phi}|}}\left(\frac{e^{-2i\mu(t-t_n)}}{\left(1 + i\frac{g|\dot{\phi}|}{\pi\mu H}A_n\right)^2} + c.c. \right)\right]
\end{align}
Similarly the result for the boson case is
\begin{equation}
	\label{equation:source:boson:expectation:value}
	\langle\chi^2_{{\bf k}_T}(\eta_1)\rangle \simeq \frac{a^{-3}(\eta_1)}{M(\eta_1)}\delta({\bf k}_T)a_n^3 \bar{n}_{\chi}\left[1 - 2e^{\frac{\pi\tilde{\mu}^2}{2g|\dot{\phi}|}}\left(\frac{e^{-2i\mu(t-t_n)}}{\left(1 + i\frac{g|\dot{\phi}|}{\pi\mu H}A_n\right)^2} + c.c.\right)\right]
\end{equation}
From (\ref{equation:source:fermion:expectation:value}) and (\ref{equation:source:boson:expectation:value}) it is already clear that the fermion order $\beta$ contribution is suppressed by an extra factor of $\frac{a_n}{a}\frac{(g|\dot{\phi}|)^{1/2}}{\mu}$. 

To proceed we will focus on case (b) of the mass function which produces novel searching templates. For the case of $N=1$, it is clear that the following relation holds
\begin{equation}
	\label{equation:mass:approximation:generalN}
	\frac{\delta^N M}{\delta\phi^N} \simeq \frac{1}{2M}\frac{\delta^N M^2}{\delta\phi^N} \simeq \frac{g^2 f^{2-N}}{\mu}\cos\left(\frac{\phi}{f} + \theta_N\right),
\end{equation}
in the parameter regime outlined in \cite{Flauger:2016idt}, where $\theta_N$ is a multiple of $\pi/2$ that gives the right trigonometric function. We can prove this result for general $N$ using mathematical induction. Starting from (\ref{equation:mass:approximation:generalN}), the $(N+1)$th derivative is
\begin{equation}
	\frac{\delta^{N+1}M}{\delta\phi^{N+1}} \simeq -\frac{g^2 f^{2-N}\cos\left(\frac{\phi}{f} + \theta_N\right)}{M^2}\frac{-g^2f\sin\left(\frac{\phi}{f}\right)}{M} - \frac{1}{M}g^2 f^{1-N}\sin\left(\frac{\phi}{f} + \theta_N\right).
\end{equation}
The first term is smaller than the second by a factor of $g^2f^2/\mu^2$ so the left hand side is approximately $\frac{1}{M}g^2f^{1-N}\cos\left(\frac{\phi}{f}+\theta_{N+1}\right)$ which is also approximately 
\begin{equation}
	\frac{1}{2M}\frac{\delta^{N+1} M^2}{\delta\phi^{N+1}} \simeq \frac{g^2 f^{1-N}}{\mu}\cos\left(\frac{\phi}{f}+\theta_{N+1}\right).
\end{equation}
This proves (\ref{equation:mass:approximation:generalN}) holds for general $N$. Combining the result above we can compute the saddle point approximation for the contributions (\ref{equation:order:beta:fermion:interm}) and (\ref{equation:order:beta:boson:interm}). The result for the fermion case is 
\begin{align}
	\label{equation:contrib:single:vertex:fermion}
& \frac{H^N}{2^{N-2}k_1^3\cdots k_N^3}\delta({\bf k}_T)\frac{\bar{n}_{\chi}}{H^3}\frac{g^2 f}{\mu}\left(\frac{H}{f}\right)^{N-1}\sum_{n=n_{\mathrm{min}}}^{n_{\mathrm{max}}} \frac{1}{-\eta_n^3}\Bigg\{ \nonumber \\%
	&-\sqrt{\frac{\pi}{2\alpha}}\left[i\left(1-i\frac{k_1}{k_T}\alpha\right) \cdots \left(1-i\frac{k_N}{k_T}\alpha\right)e^{i\alpha -i\alpha\log\frac{\alpha}{-k_T\eta_n}-i\varphi_N} + c.c.\right] \nonumber \\%
	& + \frac{2}{\pi}e^{\frac{\pi\tilde{\mu}^2}{2g|\dot{\phi}|}}\frac{(g|\dot{\phi}|)^{1/2}}{\mu} \sum_{s=\pm}\frac{\sqrt{2\pi\rho_s}}{-k_T\eta_n} \left[\frac{i\left(1-i\frac{k_1}{k_T}\rho_s \right)\cdots \left(1-i\frac{k_N}{k_T}\rho_s\right)}{\left(1-i\frac{g|\dot{\phi}|}{\pi\mu H}A_n(\hat{\eta}_s)\right)^2}e^{i\rho_s - i\rho_s\log\frac{\rho_s}{-k_T\eta_n} - i\varphi_N} + c.c.\right]
\Bigg\},
\end{align}
where we have defined the following quantities
\begin{equation}
	\alpha = \frac{\omega}{H}, ~ \rho_{\pm} = \frac{2\mu\pm \omega}{H}, ~\varphi_N = \frac{\pi}{4} - \theta_N, ~-k_T\hat{\eta}_{\pm} = \rho_{\pm},
\end{equation}
The summation of $n$ starts from the minimum $n$ (or the latest $\eta_n$) that satisfies $\eta_{n_{\mathrm{min}}} < \hat{\eta}_{\pm}$ and ends with the first/earliest production event with $n_{\mathrm{max}}$ \footnote{A different definition of $n$ is used in Section \ref{section:wkb:bogoliubov} where the first production event is labeled as $n = 1$. The definition is slightly changed here for notational simplicity and the meaning of $n$ should be clear within its context.}. It is also clear from (\ref{equation:contrib:single:vertex:fermion}) that this result respects the approximate discrete shift symmetry $\log k_T \to \log k_T + 2\pi H/\omega$ given that the shape $(k_1/k_T, \cdots, k_N/k_T)$ does not change, since $k_T$ only appears with $\eta_n$ as
\begin{equation}
	k_T \eta_n = -\frac{1}{H}\exp\left[\log k_T + \frac{2\pi H}{w}\left(n + \frac{\gamma}{2\pi}\right)\right].
\end{equation}
Such a shift can be absorbed into a relabeling of $n$.
Similarly the result for the boson case is
\begin{align}
	\label{equation:contrib:single:vertex:boson}
& \frac{H^N}{2^{N}k_1^3\cdots k_N^3}\delta({\bf k}_T)\frac{\bar{n}_{\chi}}{H^3}\frac{g^2 f}{\mu}\left(\frac{H}{f}\right)^{N-1}\sum_n \frac{1}{-\eta_n^3}\Bigg\{ \nonumber \\%
&-\sqrt{\frac{\pi}{2\alpha}}\left[i\left(1 - i\frac{k_1}{k_T}\alpha\right)\cdots \left(1-i\frac{k_N}{k_T}\alpha\right)e^{i\alpha-i\alpha\log\frac{\alpha}{-k_T\eta_n} - i\varphi_N} + c.c.\right] \nonumber \\%
& + 2e^{\frac{\pi\tilde{\mu}^2}{2g|\dot{\phi}|}}\sum_{s=\pm}\sqrt{\frac{\pi}{2\rho_s}}\left[\frac{i\left(1-i\frac{k_1}{k_T}\rho_s\right)\cdots\left(1-i\frac{k_N}{k_T}\rho_s\right)}{(1-i\frac{g|\dot{\phi}|}{\pi\mu H}A_n(\hat{\eta}_{s}))^2}e^{i\rho_s - i\rho_s\log\frac{\rho_s}{-k_T\eta_n} - i\varphi_N} + c.c.\right]
\Bigg\}.
\end{align}
A comparison between (\ref{equation:contrib:single:vertex:fermion}) and (\ref{equation:contrib:single:vertex:boson}) shows that besides the usual suppression comming from $|k_T \eta_{n_{\mathrm{min}}}|^{-3} \lesssim \left(\frac{2\mu\pm \omega}{H}\right)^{-3}$, there is an extra suppression factor of $(g|\dot{\phi}|)^{1/2}/\mu$ for the order $\beta$ contributions in the fermion case. The result of the fermion case is also further suppressed by a factor of
\begin{equation}
	\frac{2}{\pi}\frac{\sqrt{2\pi \rho_s}}{-k_T \eta_n}\cdot\frac{1}{2}\sqrt{\frac{2\rho_s}{\pi}} = \frac{2}{\pi}\cdot\frac{\rho_s}{-k_T \eta_n} = \frac{2}{\pi}\cdot\frac{-\hat{\eta}_{s}}{-\eta_n}
\end{equation}
for each production event $n$ in the summation. Since $-\eta_{n}$ gets exponentially larger than $-\eta_{n_{\mathrm{min}}}$ (which is greater than $-\hat{\eta}_s$) as $n$ increases, the contributions from the fermion case can be much smaller than those from the boson case, especially those generated by the early production events.

\section{Summary and discussion}
\label{section:summary:discussion}
In this paper we worked out in detail the contributions to the inflaton $N$ point functions from the fermion production events and compared them to those from the boson counterparts. The expectation value of the fermion source operator $\bar{\chi}\chi$ and the boson source operator $\chi^2$ have the same classical part, which is of order $|\beta|^2$, while for the order $|\beta|$ part that represents quantum interference, the fermion case is relatively suppressed by an extra factor of $k'/aM$. It is argued in section \ref{section:powerspectrum:nongaussianity} that at the level of saddle point approximation the only order $|\beta|$ contributions to inflaton $N$ point functions come from those using one $N+2$ point interaction vertex. At order $|\beta|^2$ the contributions of the two types of particle are the same. A detailed evaluation at order $|\beta|$ then shows that besides the usual suppression factor $\left(\frac{2\mu\pm\omega}{H}\right)^{-3}$, the fermion contributions as compared to those of boson are further suppressed by a factor of $(g|\dot{\phi}|)^{1/2}/\mu$ and another coming from the separation between the production time and the resonant saddle point. This distinction between fermion and boson contributions leads to the interesting possibility of getting better understandings of what type of particles and how they interact with inflaton by fitting CMB data with searching templates from these models. Similarly, this could also hint on more difficult versions of this type of models, which consider the non-adiabatic production of strings. We leave these interesting topics to future works.

The successive production of particles is treated approximately as independent from each other in this paper. In regimes where the density of the produced particles is small and quickly gets diluted by inflation this is a valid approximation. In principle, however, there can be other situations where the density is no longer small and the effect of successive productions on each other is not negligible. For the boson case this is not too hard to solve since its wave equation can be solved exactly and the recursive Bogoliubov transformation can be obtained in a similar fashion as  in \cite{Peloso:2000hy, Chung:1999ve}. The fermion case, on the other hand, seems to be less straightforward. First the wave equation (\ref{equation:wave:expanded}) has an extra imaginary part which makes it much more difficult to solve analytically.  Also the fermion production is restricted by the Pauli exclusion principle while in the boson case the production can be enhanced by previously produced particles without limit. In the regime where $|\beta|$ is not parametrically smaller than 1, the Bogoliubov coefficients need to be solved recursively as
\begin{equation}
	\begin{pmatrix}
		\alpha_n\\
		\beta_n
	\end{pmatrix}
	= T_n
	\begin{pmatrix}
		\alpha_{n-1}\\
		\beta_{n-1}
	\end{pmatrix}
\end{equation}
with the initial condition $\alpha_0 = 1, \beta_0 = 0$. $T_n$ is the transformation matrix for the $n$th production event that ensures $|\alpha|^2 + |\beta|^2 = 1$. It would be interesting to explore in this direction in future works, either analytically or numerically.

For general $N$ point functions it seems rather difficult, if possible, to derive generic close form formulae for the contributions from arbitrary interaction vertex profiles. In this work we found simple close form solutions for two extreme cases, where in one case the contributions are from $N$ three-point vertices and in the other from one $(N+2)$-point vertex. For the intermediate cases one needs to consider various possible profiles and potentially combine the methods used for the two extreme cases. Even though there is a well-defined window of parameters where at order $|\beta|^2$, contributions from $N$ three-point vertices dominate over those from other profiles, it would still be interesting to study the shapes and the relative scales of those other contributions \cite{future:work}.

\section*{acknowledgments}
I would like to thank Eva Silverstein for extensive discussion and useful comments on a draft. I am also grateful to Moritz M\"unchmeyer for initial collaboration. I am supported in part by the National Science Foundation under grant PHY-0756174 and NSF PHY11-25915 and by the Department of Energy under contract DE-AC03-76SF00515.
	
\begin{appendices}
		\section{More on Bogoliubov coefficients} \label{appendix:sec:bogo}
		In this section we explore the possibility of deriving the exact recurrence relation between the coefficients $A_n$, $B_n$ after the $n$th production event and those before it. As noted in section \ref{subsection:wkb:bogo}, the wave equation (\ref{equation:wave:expanded}) can be greatly simplified in the $|\eta| \to \infty$ limit, which is where the matching of WKB solutions happens. One can use the asymptotic behavior of the solution to (\ref{equation:wave:expanded}) in this region to match with the two WKB solutions. Similar to the rotation method used in the main text, the same problem regarding the branch cut of $M$ arises with this method. In the $|\eta|\to\infty$ limit, the original wave equation (\ref{equation:wave:expanded}) simplifies to 
		\begin{equation}
			\label{equation:wave:simplified}
			u_+'' + \left(k^2 + a_n^2 \tilde{\mu}^2 + g^2 a_n^4\dot{\phi}^2 \eta ^2 \pm iga_n^2|\dot{\phi}|
			\right)u_+ = 0,
		\end{equation}
		where the sign for the imaginary term depends on which branch of $M$ is being considered. In order to make use of the solution to (\ref{equation:wave:simplified}) for matching, it needs to be on the same branch which covers both of the $\eta \to -\infty$ and $\eta \to +\infty$ limits since otherwise one would have two different approximate differential equations, one with a positive imaginary term and another with a negative one. The solutions then would have to be connected via the small $|\eta|$ region where there is no branch cut. Connecting the two branches in this way is as hard as solving (\ref{equation:wave:expanded}) exactly. Therefore we pick the same branch cut as shown in Figure \ref{fig:mass:phase:diagram:right} so that the imaginary part of (\ref{equation:wave:simplified}) always has a negative sign.
		
	With this definition, the solution to (\ref{equation:wave:simplified}) is 
	\begin{equation}
		\label{equation:uplus:sol}
		u_+(\eta) \approx C_1 D_{p_1}\left((1+i)\nu_1\eta\right) + C_2 D_{p_2}\left((-1+i)\nu_2\eta\right),
	\end{equation}
	where $C_1$, $C_2$ are two constants and $D_p(z)$ is the parabolic cylinder function. $p_1$ and $p_2$ are defined as 
	\begin{equation}
		p_1 = -\frac{i\nu_0^2}{2\nu_1^2}, \quad p_2 = -1 + \frac{i\nu_0^2}{2\nu_1^2},
	\end{equation}
	while all the other symbols are defined in the main text. In the meantime, the WKB solution is given in (\ref{equation:WKB:uplus}). Note again that there is only one set of $C_1$ and $C_2$ for both the $\eta \to -\infty$ and $\eta \to +\infty$ limit while $A$, $B$ are really $A_{n-1}$, $B_{n-1}$ before the $n$th production event and $A_n$, $B_n$ afterwards. The asymptotic behavior of (\ref{equation:uplus:sol}) in the $\eta \to -\infty$ limit is \cite{bender:orszarg}
	\begin{align}
		\label{equation:u0:expansion:neg}
		u_0 \sim & C_1 e^{-\frac{i}{2}|x|^2}\left(\sqrt{2}e^{-i\frac{3\pi}{4}}|x|\right)^{-\frac{i\nu_0^2}{2\nu_1^2}}  \nonumber \\%
			& +e^{\frac{i}{2}|x|^2}|x|^{-1 + \frac{i\nu_0^2}{2\nu_1^2}}\left[
				C_2\left(\sqrt{2}e^{-i\frac{\pi}{4}}\right)^{-1 + \frac{i\nu_0^2}{2\nu_1^2}} -
					C_1 i \frac{\sqrt{2\pi}}{\Gamma\left(\frac{i\nu_0^2}{2\nu_1^2}\right)}
						e^{-i\pi\left(\frac{1}{2}-\frac{i\nu_0^2}{2\nu_1^2}\right)}
						\left(\sqrt{2}e^{-i\frac{3\pi}{4}}\right)^{-1+\frac{i\nu_0^2}{2\nu_1^2}}\right],
	\end{align}
	and in the $\eta \to +\infty$  limit
	\begin{align}
		\label{equation:u0:expansion:pos}
		u_0 \sim & e^{-\frac{i}{2}|x|^2} |x|^{-\frac{i\nu_0^2}{2\nu_1^2}}
			\left[C_1\left(\sqrt{2} e^{i\frac{\pi}{4}}\right)^{-\frac{i\nu_0^2}{2\nu_1^2}} + 
				C_2 i \frac{\sqrt{2\pi}}{\Gamma\left(1 - \frac{i\nu_0^2}{2\nu_1^2}\right)}
					e^{i\pi\left(-\frac{1}{2} + \frac{i\nu_0^2}{2\nu_1^2}\right)}
					\left(\sqrt{2}e^{i\frac{3\pi}{4}}\right)^{-\frac{i\nu_0^2}{2\nu_1^2}}\right]   \nonumber \\%
			& + e^{\frac{i}{2}|x|^2}|x|^{-1 + \frac{i\nu_0^2}{2\nu_1^2}} C_2\left(\sqrt{2} e^{i\frac{3\pi}{4}}\right)^{-1 + \frac{i\nu_0^2}{2\nu_1^2}},
	\end{align}
	where ``$\sim$'' denotes ``asymptotic to''. They need to match with those of the WKB solutions, which in the $\eta\to-\infty$ limit is
	\begin{align}
		\label{equation:WKB:expansion:neg}
		u \sim & e^{-\frac{i}{2}|x|^2}|x|^{-\frac{i\nu_0^2}{2\nu_1^2}}\left(\frac{2\nu_1}{\nu_0}\right)^{-\frac{i\nu_0^2}{2\nu_1^2}}\sqrt{2}B_{n-1} e^{-\frac{i\nu_0^2}{4\nu_1^2}} 
			+ e^{\frac{i}{2}|x^2|}|x|^{-1+\frac{i\nu_0^2}{2\nu_1^2}}\left(\frac{2\nu_1}{\nu_0}\right)^{\frac{i\nu_0^2}{2\nu_1^2}} A_{n-1} \frac{k}{\sqrt{2} \nu_1}e^{\frac{i\nu_0^2}{4\nu_1^2}},
	\end{align}
	and in the $\eta \to +\infty$
	\begin{align}
		\label{equation:WKB:expansion:pos}
		u \sim & e^{\frac{i}{2}|x|^2}|x|^{-1 + \frac{i\nu_0^2}{2\nu_1^2}}\left(\frac{2\nu_1}{\nu_0}\right)^{\frac{i\nu_0^2}{2\nu_1^2}}B_n \frac{k}{\sqrt{2}\nu_1}e^{\frac{i\nu_0^2}{4\nu_1^2}} 
		 	+ e^{-\frac{i}{2}|x|^2} |x|^{-\frac{i\nu_0^2}{2\nu_1^2}} \left(\frac{2\nu_1}{\nu_0}\right)^{-\frac{i\nu_0^2}{2\nu_1^2}} A_n \sqrt{2} e^{-\frac{i\nu_0^2}{4\nu_1^2}}.
	\end{align}
	After some algebra, this matching gives the following recurrence relation
	\begin{align}
		\label{equation:A_n}
		A_n &= B_{n-1} e^{-\frac{\pi\nu_0^2}{2\nu_1^2}} + A_{n-1} e^{\frac{i\pi}{4}} 2\sqrt{\pi} e^{-\frac{\pi\nu_0^2}{4\nu_1^2}}\frac{\left(\frac{\nu_0^2}{2e\nu_1^2}\right)^{-\frac{i\nu_0^2}{2\nu_1^2}}}{\Gamma\left(-\frac{i\nu_0^2}{2\nu_1^2}\right)}\frac{\nu_1 k}{\nu_0^2},\\%
		\label{equation:B_n}
		B_n &= -A_{n-1} e^{-\frac{\pi\nu_0^2}{2\nu_1^2}} + 
			B_{n-1} e^{-\frac{i\pi}{4}}2\sqrt{\pi} e^{-\frac{\pi\nu_0^2}{4\nu_1^2}}\frac{\left(\frac{\nu_0^2}{2e\nu_1^2}\right)^{\frac{i\nu_0^2}{2\nu_1^2}}}{\Gamma\left(\frac{i\nu_0^2}{2\nu_1^2}\right)}\frac{\nu_1}{k}.
	\end{align}
		
	As noted in section \ref{subsection:wkb:bogo} $A$ and $B$ satisfy the normalization $|A|^2 + |B|^2 = 1$. This serves as a good consistency check of the above result. A straightforward calculation yields
	\begin{align}
		\label{equation:normalization:check}
		|A_n|^2 + |B_n|^2 = &  |A_{n-1}|^2 \left[1 + \left(\frac{k^2}{\nu_0^2} - 1\right)\left(1 - e^{-\frac{\pi\nu_0^2}{\nu_1^2}}\right)\right] \nonumber \\%
			&+ |B_{n-1}|^2\left[1 + \left(\frac{\nu_0^2}{k^2} - 1\right)\left(1 - e^{-\frac{\pi\nu_0^2}{\nu_1^2}}\right)\right] \nonumber \\%
			&+ 2\mathrm{Re}\left[A_{n-1}^* B_{n-1} e^{-\frac{i\pi}{4}}\frac{\left(\frac{\nu_0^2}{2e\nu_1^2}\right)^{\frac{i\nu_0^2}{2\nu_1^2}}}{\Gamma\left(\frac{i\nu_0^2}{2\nu_1^2}\right)}\right]2\sqrt{\pi}e^{-\frac{3\pi\nu_0^2}{4\nu_1^2}}\frac{\nu_1}{k}\left(\frac{k^2}{\nu_0^2} - 1\right).
	\end{align}
	This value is not exactly 1 given $|A_{n-1}|^2 + |B_{n-1}|^2 = 1$. The reason is that the recurrence relation (\ref{equation:A_n}) and (\ref{equation:B_n}) are obtained by matching the WKB solutions with (\ref{equation:uplus:sol}), which is the solution to the approximate differential equation obtained under the assumption $\tilde{\mu}^2 \ll g^2 a_n^2 \dot{\phi}^2 \eta^2$  and $g^2 a_n^4\dot{\phi}^2\eta^2 \simeq \nu_0^2 = k^2 + a_n^2\tilde{\mu}^2$. The two approximations imply that $k^2 \simeq \nu_0^2$, which along with  (\ref{equation:normalization:check}) leads to $|A_n|^2 + |B_n|^2 \simeq 1$. Therefore at this level of approximation, the result (\ref{equation:normalization:check}) is consistent with the normalization condition. We may understand the failure of obtaining the recurrence relation of coefficients $A_n$ and $B_n$ using this approximation in the following way.  The asymptotic behavior of the exact solution to the original differential equation (\ref{equation:wave:expanded}) should really be
	\begin{equation}
		\label{equation:uplus:sol:exact:asymp}
		u_+(\eta) \approx \tilde{C}_1(\eta) D_{p_1}\left((1+i)\nu_1\eta\right) + \tilde{C}_2(\eta) D_{p_2}\left((-1+i)\nu_2\eta\right),
	\end{equation}
	where now $\tilde{C}_1(\eta)$ and $\tilde{C}_2(\eta)$ are functions of $\eta$ and they are asymptotically constants. The solution to the approximated differential equation (\ref{equation:uplus:sol}) assumes that the constants are universal on the lower half plane. However $\tilde{C}_1(\eta)$ and $\tilde{C}_2(\eta)$ can have more complicated Stokes phenomenon and have different constant values in different sections. We expect that this omission of Stokes phenomenon to be the reason of failing to recover the recurrence relation of $A_n$ and $B_n$'s. 

\end{appendices}

\end{document}